\documentclass[12pt]{article}
\usepackage{amsmath,amsfonts,amssymb,amsthm,amstext,amscd,eucal}
\usepackage{hyperref}
\usepackage{epsfig}\usepackage{graphicx}
\usepackage[all]{xy}

\makeatletter \@addtoreset{equation}{section}

\makeatletter\renewcommand\section{\@startsection {section}{1}{\z@}%
                                   {-3.5ex \@plus -1ex \@minus -.2ex}
                                   {2.3ex \@plus.2ex}%
                                   {\normalfont\large\bfseries}}
\renewcommand\subsection{\@startsection{subsection}{2}{\z@}%
                                     {-3.25ex\@plus -1ex \@minus -.2ex}%
                                     {1.5ex \@plus .2ex}%
                                     {\normalfont\bfseries}}

\newcommand{\be}{\begin{equation}}
\newcommand{\ee}{\end{equation}}
\newcommand{\beq}{\begin{eqnarray}}
\newcommand{\eeq}{\end{eqnarray}}
\newcommand{\bea}{\begin{eqnarray}}
\newcommand{\eea}{\end{eqnarray}}
\newcommand{\bse}{\begin{subequations}}
\newcommand{\ese}{\end{subequations}}

\def\[{\left [}
\def\]{\right ]}
\def\({\left (}
\def\){\right )}

\def\cN{{\cal N}}

\def\r2{\sqrt{2}}

\def\V4{{\cal V}_{\CM_4}}

\newcommand{\U}{\mathrm{U}}

\newcommand{\SU}{\mathrm{SU}}

\newcommand{\nn}{\nonumber}

\newcommand{\ie}{{\em i.e.}\ }

\newcommand{\ads}[1]{{\rm AdS}_{#1}}

\def\sst#1{{\scriptscriptstyle #1}}

\def\1{{\sst{(1)}}}

\def\CM{{\cal M}}
\def\CN{{\cal N}}
\def\CO{{\cal O}}

\renewcommand{\ads}[1]{${\rm AdS}_{#1}$}

\newcommand{\bbibitem}[1]{\bibitem{#1}\marginpar{#1}}

\def\Label#1{\label{#1}%
  \smash{\hbox to0pt{\raise1ex\hbox{\tiny[#1]}\hss}}}
\def\noLabels{\let\Label=\label}
\def\nobbibitem{\let\bbibitem=\bibitem}

\begin{document}

\noLabels 
\nobbibitem 

\begin{titlepage}

\begin{flushright}\vspace{-3cm}
{\small
IPM/P-2011/054 \\
\today }\end{flushright}
\vspace{-.5cm}

\begin{center}
\centerline{{\Large{\bf{Emergent IR dual 2d CFTs in charged AdS$_5$ black holes }}}} \vspace{4mm}

{\large{{\bf Jan de Boer,\footnote{e-mail: J.deBoer@uva.nl}$^{,ab}$ Maria Johnstone,\footnote{e-mail:
M.J.F.Johnstone@sms.ed.ac.uk}$^{,c}$ M.M. Sheikh-Jabbari\footnote{e-mail:
jabbari@theory.ipm.ac.ir}$^{,d}$ \\
and Joan Sim\'on\footnote{e-mail:
j.simon@ed.ac.uk}$^{,c}$ }}}
\\

\vspace{5mm}

\bigskip\medskip
\begin{center}
{$^a$ \it Instituut voor Theoretische Fysica,
Science Park 904\\
Postbus 94485,
1090 GL Amsterdam,
The Netherlands}\\
\smallskip
{$^b$ \it Gravitation and AstroParticle Physics Amsterdam}\\
\smallskip
{$^c$ \it School of Mathematics and Maxwell Institute for Mathematical Sciences,\\
King's Buildings, Edinburgh EH9 3JZ, United Kingdom}\\
\smallskip
{$^d$ \it School of Physics, Institute for Research in Fundamental
Sciences (IPM),\\ P.O.Box 19395-5531, Tehran, Iran}\\
\smallskip
\end{center}
\vfil

\end{center}
\setcounter{footnote}{0}

\begin{abstract}
\noindent

We study the possible dynamical emergence of IR conformal invariance describing the low energy excitations of
near-extremal R-charged global AdS${}_5$ black holes. We find interesting behavior especially when we tune parameters
in such a way that the relevant extremal black holes have classically vanishing horizon area, i.e. no classical ground-state
entropy, and when we combine the low energy limit with a large $N$ limit of the dual gauge theory.
We consider both near-BPS and non-BPS regimes and their near horizon limits, emphasize the differences between the local AdS${}_3$ throats emerging in either case, and
discuss  potential dual IR 2d CFTs for each case. We compare our results with the predictions obtained from the Kerr/CFT correspondence,
and obtain a natural quantization for the central charge of the near-BPS emergent IR CFT which we interpret in terms of
the open strings stretched between giant gravitons.

\end{abstract}

\end{titlepage}
\renewcommand{\baselinestretch}{1.05}  
\tableofcontents


\section{Introduction}

The microscopic understanding of non-extremal black holes is an important problem in theoretical physics. Their universal Rindler near horizon geometries and the existence of chiral Virasoro algebras generated by diffeomorphisms preserving this structure raises the possibility of having a conformal field theory (CFT) description for these systems \cite{Carlip:1998wz}\footnote{See \cite{Carlip:2011ax} for a more recent discussion and \cite{Simon:2011zz} for a review of these ideas in a more general holographic context.}, generalising the structure uncovered in AdS${}_3$ \cite{Brown-Henneaux,andy-btz}.

Progress was recently achieved by pursuing these ideas for finite extremal black holes whose near horizon geometry includes an AdS${}_2$ factor\footnote{This is a theorem in d=4,5 dimensions, and it extends to higher dimensions, under some isometry
assumptions \cite{harvey}. The theorem also allows global AdS${}_3$ geometries for a class of horizons generated by  static null Killing vectors. This possibility is different from the one we will discuss in this note.}. In \cite{Kerr-CFT,Hartman:2008pb}, it was shown that one can semiclassically associate a chiral Virasoro algebra to the near-horizon  geometry of finite extremal black holes,
using asymptotic symmetry group considerations \cite{Barnich:2001jy}. The existence of a dual chiral CFT accounting for the black hole entropy was also conjectured.

Despite the success in computing the entropy of extremal black holes, either using the Kerr/CFT correspondence \cite{Kerr-CFT,Hartman:2008pb} or the entropy function formalism based on the enhancement of symmetry of their near horizons \cite{senpapers}, there are arguments reviewed in section \ref{sec:gphil} suggesting these AdS${}_2$ geometries do not generically represent a decoupled conformal field theory. Even if they would, the AdS/CFT machinery would suggest these theories may be dynamically trivial \cite{Sen-AdS2/CFT1,us,EVH-BTZ}, in the sense that they only contain  degeneracy of the vacuum in their spectrum.

\begin{figure}
\centerline{\epsfig{file=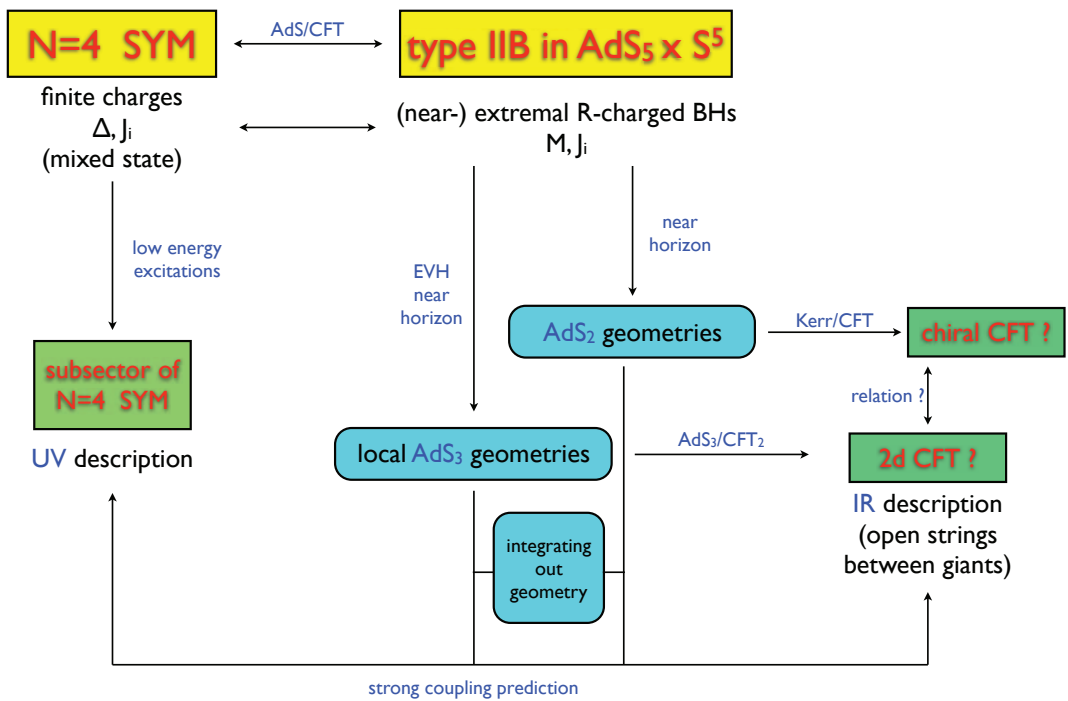,width=15cm}}
\caption{Embedding our set-up in an UV CFT using an extremal asymptotically AdS black hole, studying its low energy excitations through its near horizon geometry and identifying potential IR 2d CFTs describing them. \label{fig3}}
\end{figure}

In this note, we want to understand the circumstances under which the near horizon limit of extremal black holes exhibit non-trivial  dynamics and analyse the emergence of low energy dynamical conformal symmetry. In our analysis we consider certain near-extremal static R-charged AdS$_5$ black holes and their near horizon geometries to study their low energy excitations (see Figure \ref{fig3} to illustrate our set-up). We focus on black holes whose extremal limit has a vanishing horizon area (in units of AdS$_5$ radius). They belong to the family of Extremal Vanishing Horizon (EVH) black holes whose near horizon geometry develops a local AdS$_3$-like throat, signalling the possible existence of an (IR) dual 2d CFT which captures the low energy dynamics around the background of EVH black holes. By assumption, these asymptotic AdS$_5$ black holes have a dual ultraviolet (UV) CFT description, as a thermal mixed state, using the standard AdS/CFT correspondence \cite{witten-phase,ads-cft}. In the near BPS case, they can even be microscopically interpreted as distributions of smeared giant gravitons \cite{Myers:2001aq}. If this UV CFT is non-singular and defined on a compact space, which is true for our case where the UV CFT is ${\cal N}=4$ SYM on $R\times S^3$, its spectrum is gapped and at low enough energies above the EVH black hole, no dynamics should be left.

To circumvent this conclusion we will take {\it large central charge} limits of this UV CFT, keeping the AdS radius fixed, i.e. large $N$ limits. Besides the well known possibility involving planar black holes, briefly reviewed in section \ref{sect5}, one can also consider vanishing horizon black holes keeping the near-extremal entropy, or rather its density, fixed. We are primarily interested in understanding whether the emergent local AdS${}_3$ geometries appearing in these cases describe the low energy excitations of the original UV CFT in terms of
an infra-red (IR) 2d CFT.

In section \ref{sect2}, we identify two distinct regimes where to study this phenomena: a near-BPS and a non-BPS regime.
In sections \ref{sect:nearbps} and \ref{sec:nonbps}, we study these by taking different near horizon limits. We discuss the important geometrical differences between the two, identify the relevant sectors of ${\cal N}=4$ SYM in each case and compute the standard CFT${}_2$ parameters which we compare with Kerr/CFT predictions in section \ref{sec:kerr-cft}. Further advantages of our approach are the natural quantisation of the central charges emerging in the near-BPS regime and the potential BMN-like \cite{BMN} interpretation that our near horizon limits offer. In either case, we attempt to provide an interpretation for our results and comment on the importance/limitations of taking the near horizon limit in our summary and outlook. In the Appendix we discuss near horizon limit of EVH black holes in the family of Myers-Perry black holes \cite{myersperry}.

\section{General philosophy}
\label{sec:gphil}

A generic asymptotically AdS$_{d+1}$ black hole is described by a thermal mixed state in the dual UV CFT theory. Excitations above it will generically have a gap if the latter is defined on $\mathbb R\times$S${}^{d-1}$ and is non-singular. Thus, probing the system at sufficiently low energies above the black hole, but below the gap, one expects to keep the degeneracy of the ground state of extremal black hole (black hole entropy), but no non-trivial dynamics.

This argument suggests that if there is any emergent CFT in the deep IR, associated with the near horizon geometry, the latter will contain no non-trivial dynamics. In particular, for generic extremal black holes, whose near horizon geometries include AdS${}_2$ throats, we would conclude that such IR CFTs would only contain the vacuum state and its degeneracy. This seems to be consistent with arguments such as AdS${}_2$-fragmentation \cite{fragmentation}, AdS${}_2$/CFT${}_1$ considerations  \cite{Sen-AdS2/CFT1} or the absence of gravitational perturbations preserving the near horizon of 4d extremal Kerr \cite{Horowitz-Marolf-Harvey}.\footnote{This conclusion is expected to be much more subtle in a generic situation, given the existence of multi-center AdS${}_2$ configurations when the cosmological constant vanishes. As already emphasised in \cite{fragmentation}, these classical configurations survive the low energy limit. Recently, further configurations were found in different supergravity theories sharing this same feature. A proper microscopic understanding of these is not known, though they were already argued to correspond to a physical situation where the Higgs and Coulomb branches of the dual gauge theory coincide \cite{fragmentation}.}

This conclusion can be bypassed, as illustrated in Figure \ref{fig1}, if one violates one of the above assumptions:
\begin{itemize}
\item[(i)] if the UV CFT is defined on a {\it non-compact space}, its spectrum will be continuous and non-trivial physical excitations may exist at low energies. Non-compactness of the boundary theory also implies non-compactness of the black hole horizon and consequently, the vanishing of the two dimensional Newton's constant obtained from dimensional reduction of the gravity theory over the near horizon geometry of extremal black hole. (The AdS${}_2$ space appearing in the near horizon geometry is a solution to this 2d gravity theory.) The latter also bypasses the fragmentation argument. This set-up has prominently appeared  in some recent applications of the AdS/CFT correspondence to condensed matter systems \cite{AdS-CMT,Faulkner:2009wj}. Even in these cases, there is no evidence for decoupling of the UV and IR physics (though some non-analytical features are seemingly captured by the AdS${}_2$ throats).

\begin{figure}[ht]
\centerline{\epsfig{file=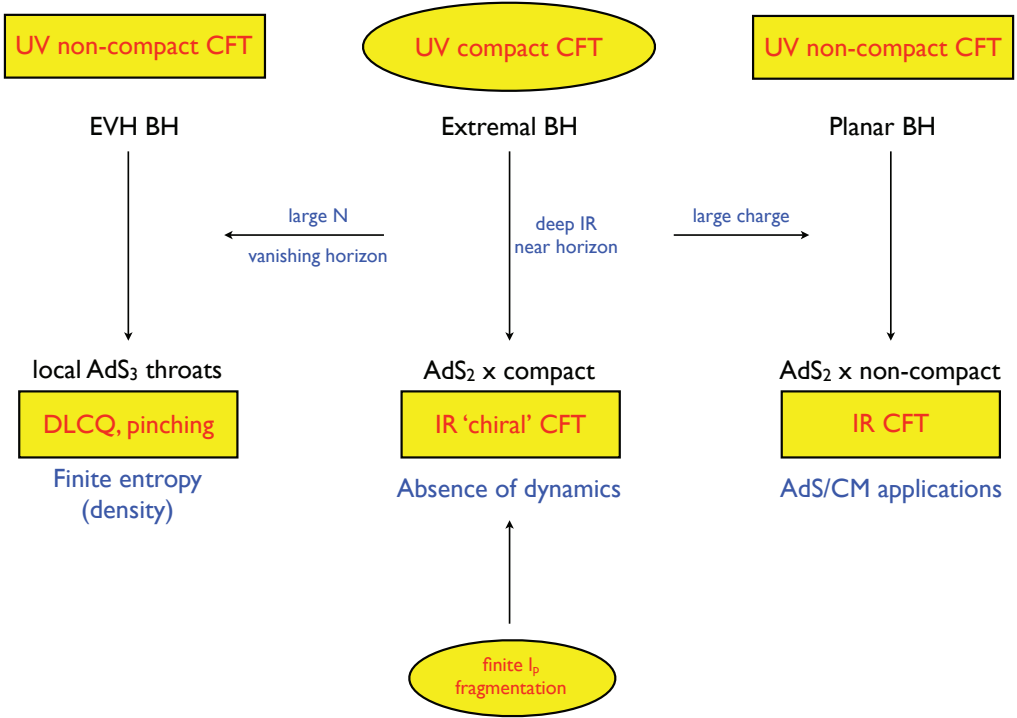,width=15cm}}
\caption{Large $N$ limits in non-singular UV CFTs to get non-trivial physical excitations above extremal black holes at low energies. \label{fig1}}
\end{figure}

\item[(ii)] if we {\it decrease the mass gap} $\delta \Delta_{\text{UV}}$ of the strongly coupled dual UV CFT.  At weak coupling, this implies a large central charge limit, both in a near BPS and far from BPS situations, in view of the gap $\delta \Delta_{\text{UV}} \sim1/c$ one obtains from the long string picture. At strong coupling and for BPS spectrum, the same conclusion was shown to hold in \cite{deBoer:2009un}, using the emergence of deep throats. As far as we know, this conclusion has not been extended to strongly coupled far from BPS situations, where the standard lore is that such spectrum will look random, so that $\delta \Delta_{\text{UV}}\sim e^{-S}$ \cite{hep-th/0604045}.
Either way, string theory realisations of these scenarios typically involve a large charge limit. For a given temperature, these would give rise to a divergent entropy. To keep the latter {\it finite}, one must combine the large charge limit with a
\emph{vanishing horizon} limit, which in turn also demands \emph{vanishing temperature} (extremal) limit of the original black hole. In the case of asymptotic AdS$_5$ EVH black holes, as we will discuss in this paper, this corresponds to a certain large $N$ limit. In both BPS and non-BPS EVH cases taking the large $N$ limit together with near extremal limit will open up the possibility of having
non-trivial excitations and dynamics.
\end{itemize}

In any statistical mechanical system in equilibrium entropy is a positive-definite function of charges and temperature and the entropy can vanish only at zero temperature, i.e. the vanishing entropy limit of any system corresponds to its low temperature (IR) expansion.\footnote{We note that the converse, the usual statement of third law of thermodynamics, does not hold in the cases involving extremal black holes; i.e. extremal black holes generically have a non-zero finite entropy.}
Therefore, we consider the low temperature IR expansion  for the gravitational (Bekenstein-Hawking) entropy of a black hole%
\begin{equation}\label{entropy-expansion}
  S(q_i, N; T)=S_0(q_i; N)+S_1(q_i;N)T+S_2(q_i;N) T^2+\cdots
\end{equation}

where $q_i$ stand for the different black hole charges and $N$ for the rank of the dual gauge group or a quantum number playing a similar role. Generic extremal black holes have non-zero $S_0(q_i;N )$, providing the dominant contribution to the entropy in this IR limit. There may be specific extremal black holes for which the coefficients $S_n$ are zero for $n<k$. In that situation, the leading contribution to the entropy is $S\sim S_k(q_i;N)T^k$ and one may speculate on the existence of a dual IR $k+1$ dimensional CFT, since $S\sim c_k T^k$ follows from conformal invariance with $c_k$ being some effective central charge.\footnote{Note that a similar low temperature expansion and similar reasoning also applies to the well established (near-BPS) black $p$-brane solutions, which for $k=2,3,5$ lead to the usual (maximally supersymmetric) AdS$_{k+2}$/CFT$_{k+1}$ examples.}
\begin{figure}[ht]
\centerline{\epsfig{file=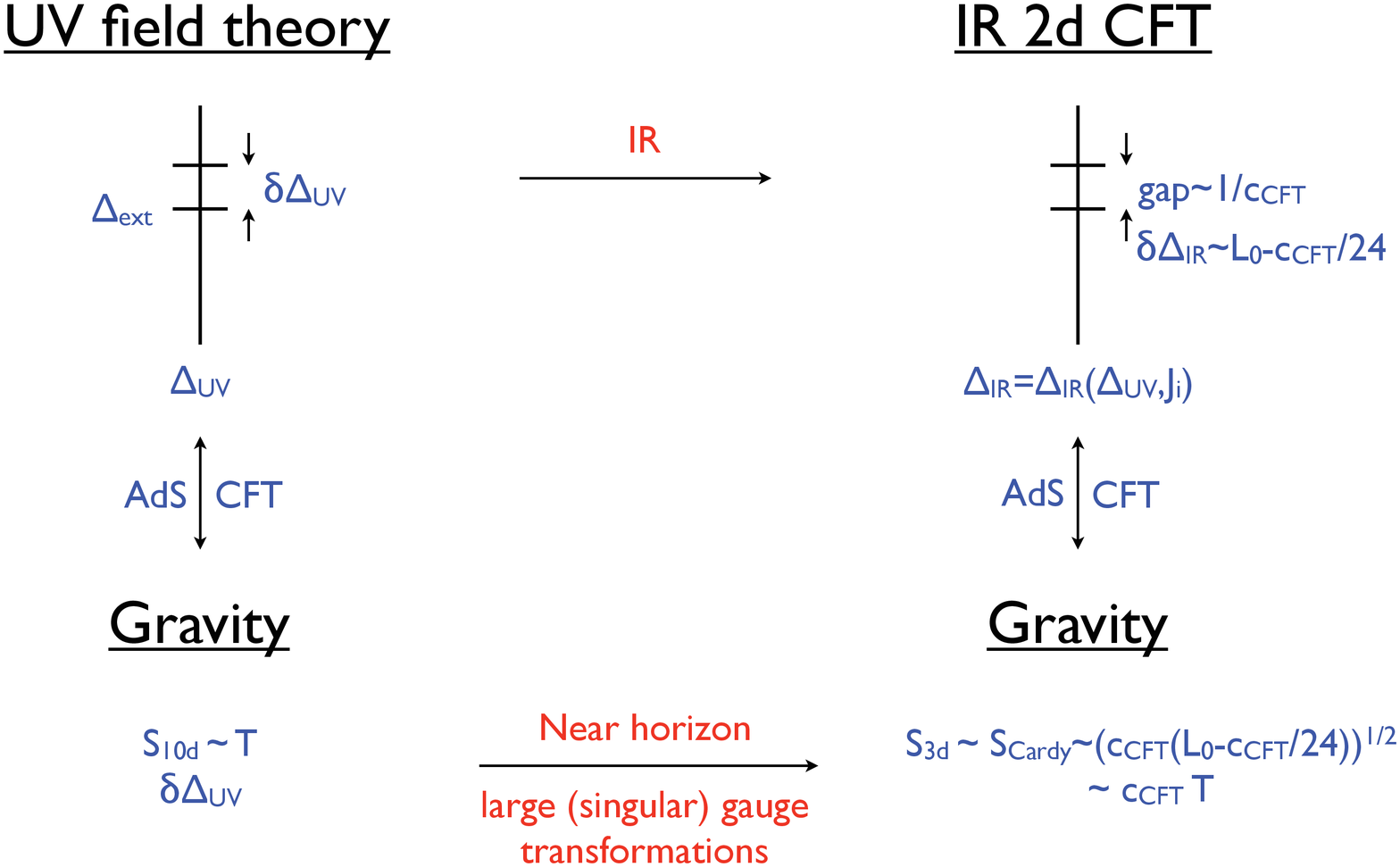,height=12cm,width=15cm}}
\caption{Set of scales and IR description of UV excitations. \label{fig2}}
\end{figure}

The possibility that such an IR CFT may be supported by a near horizon local AdS$_{k+2}$ throat has already appeared in the literature  in
the case $k=1$. The fact that the near horizon geometry of an EVH black hole has a local AdS$_3$ throat was originally pointed out in \cite{bardeenhorowitz} for extremal 5d Kerr black holes
with one vanishing angular momentum, where the near horizon geometry involves a {\it pinching} AdS$_3$ orbifold.\footnote{The word pinching AdS$_3$ orbifold was coined in \cite{EVH-BTZ}, where the simplest possible EVH black hole, namely the massless BTZ black hole, and its possible near horizon limits were discussed.  The pinching AdS$_3$ orbifold  is a singular geometry which can be thought of as AdS$_3/Z_K$ in the $K\to\infty$ limit.} As discussed in the Appendix, this statement can be easily generalised to higher dimensional Myers-Perry black holes \cite{myersperry}: extremal Myers-Perry black holes with {\it one} vanishing angular momentum develop such throats. The appearance of an AdS$_3$ throat  for R-charged AdS$_d$ (for $d=4,5$) black holes was reported in \cite{Balasubramanian:2007bs,Fareghbal:2008ar,Fareghbal:2008eh} when one of the R-charges is parametrically smaller than the rest. It was recently proved that
the near horizon geometry of any EVH solution of four dimensional Einstein-Maxwell-Dilaton theory has a pinching AdS$_3$ orbifold throat \cite{SheikhJabbaria:2011gc}.
Interestingly, in the Kerr/CFT context, attempts to give a microscopic derivation of the conjecture also ended up exploring regions in parameter space where the horizon became of zero size \cite{G-S,Compere}. These ideas were extended, under certain conditions, for extremal vanishing horizons in four and five dimensions in \cite{terashima,recent-evh, terashima-2}.

A common feature of all these $k=1$ examples giving rise to local AdS${}_3$ throats is that both horizon area and temperature of the extremal black hole tend to zero keeping their ratio finite
\begin{equation}
  A,\,T\to 0 \quad \quad {\rm with} \quad \quad \frac{A}{T}\,\,{\rm finite}
\label{def:evh}
\end{equation}
It is this ratio that suggests the potential emergence of a gravitational thermodynamical system in 3 dimensions having a 2d CFT dual with finite central charge. This is indeed the philosophy advocated in \cite{SheikhJabbaria:2011gc} giving rise to the so called EVH/CFT correspondence.

To sum up, as illustrated in Figure \ref{fig2}, one starts with a black hole in AdS whose Bekenstein-Hawking entropy, at low temperatures, satisfies
\begin{equation}
  S_{{\rm 10d}} = \frac{A}{4G_N^{(10)}} \propto T.
\label{eq:evh10d}
\end{equation}
This relation holds in a regime of black hole charges determining a specific set of UV CFT charges $\Delta_{{\rm UV}}$ and $J_i$.
The question is whether there exists an alternative description for the physics of low-lying excitations with
quantum numbers $\delta\Delta_{{\rm UV}}$ and $\delta J_i$ satisfying $\delta\Delta_{{\rm UV}}\ll \Delta_{{\rm UV}}$.
We explore the proposal that this alternative description is in terms
of a 2d CFT.

Geometrically, near-extremal horizon limits typically involve non-trivial (singular) large gauge transformations defining the near horizon geometry (IR description) in terms of the isometry coordinates of the boundary geometry (UV CFT theory)
\begin{equation}
  r=r_h + \epsilon\,\rho, \quad \quad \varphi_i^{\text{IR}}=\varphi_i^{\text{IR}}\left(\varphi_i^{\text{UV}},t^{\text{UV}},\epsilon\right), \quad \quad  t^{\text{IR}}=t^{\text{IR}}\left(\varphi_i^{\text{UV}},t^{\text{UV}},\epsilon\right)\quad \epsilon\to 0
\end{equation}
This suggests, as also expected from a purely field theoretical perspective, the existence of a non-trivial relation between  the UV and IR Hamiltonians. If the IR theory is conformal, there will therefore be an interesting relation between quantum numbers of the form
$\Delta_{{\rm IR}}=\Delta_{{\rm IR}}(\Delta_{{\rm UV}},J_i)$.\footnote{For example, in a near-BPS situation, one expects $\Delta_{{\rm IR}}=\Delta_{{\rm UV}}-\sum_i J_i$. We will explicitly see this feature emerging in section \ref{sect:nearbps}. The importance of these singular large gauge transformations for extremal black holes has been emphasised in \cite{Simon:2011zz, Hartman:2008dq}.} Notice there is no guarantee the emergent 2d CFT would be local in the original UV description. This is expected not to be the case whenever the charges involved correspond to R-charges (internal charges). It would be very interesting to develop a renormalisation group perspective  (interpreted as integrating out geometry \cite{int-out}) on these non-trivial relations.

In view of the 10 dimensional thermodynamical relation \eqref{eq:evh10d}, one is looking for a reinterpretation of the UV spectrum in terms of an effective IR 2d CFT, as illustrated in Figure \ref{fig2}, whose central charge and energy fluctuations satisfy
\begin{equation}
  c_{{\rm CFT}} \propto \frac{S_{{\rm 10d}}}{T}, \quad \quad L_0-\frac{c_{{\rm CFT}}}{24}\sim \delta\Delta_{{\rm IR}}\,.
\end{equation}
If the IR CFT has a gravitational dual, presumably related to the near horizon geometry of the initial (near-)extremal black hole,
one would in particular expect the Brown-Henneaux  relation \cite{Brown-Henneaux}
\begin{equation}
   c_{{\rm CFT}} = \frac{3\ell_3}{2G_N^{(3)}}
\end{equation}
to hold, which will provide a useful consistency check of this framework.

Notice that keeping the {\it entropy finite}, while $A\to 0$, will require us to rescale the 10d Newton's constant. In our set-up, which keeps the UV AdS radius fixed, this requires an $N\to\infty$ limit, which we will discuss in more detail in the upcoming sections.

\section{Vanishing horizon limits for R-charged AdS${}_5$ black holes}
\label{sect2}

In this section, we review the characterisation of extremal vanishing horizons among R-charged AdS${}_5$ black holes.
These are solutions of type IIB supergravity with constant dilaton, and metric and RR 4-form
potential given by \cite{Cvetic:1999xp}%
\begin{eqnarray}%
ds_{10}^2 &=& {\sqrt{\Delta}}\left(-\frac{f}{H_1H_2H_3}dt^2+\frac{dr^2}{f} +r^2\,d
\Omega_3^2\right)+\frac{L^2}{\sqrt{\Delta}}\left(\sum_{i=1}^3
H_i\left(d\mu_i^2+\mu_i^2\left[d\phi_i+a_i\,dt\right]^2\right)\right)\nonumber \\
C_4 &=& - \frac{r^4}{L} \Delta\, dt \wedge d^3\Omega - L \sum_{i=1}^3
{\tilde q}_i \,\mu_i^2 \,\left( L\, d\phi_i - \frac{q_i}{{\tilde
q}_i} dt \right) \wedge d^3\Omega\,.
\label{10-dim-general}
\end{eqnarray}
The configuration is determined by a set of scalar functions $\{H_i,\,f,\,\Delta\}$ and gauge
fields $\{a_i\}$
\bse\label{Hif-10d}
\begin{align}
 H_i = 1 + \frac{q_i}{r^2}, &\qquad a_i = \frac{{\tilde q}_i}{q_i} \frac{1}{L}
 \left(
\frac{1}{H_i}-1 \right), \,\,\,i=1,2,3 \\
f = 1 - \frac{\mu}{r^2} + \frac{r^2}{L^2} H_1 \, H_2 \, H_3, &
\qquad \Delta = H_1 \, H_2 \, H_3
\left[\frac{\mu_1^2}{H_1}+\frac{\mu_2^2}{H_2}+\frac{\mu_3^2}{H_3}\right]\,,
\end{align}\ese%
the unit radius 3-sphere metric $d\Omega_3^2$ and a further 2-sphere $\mu_1^2 + \mu_2^2+\mu_3^2=1$.

These solutions have four independent parameters $\{\mu,\,q_i\}$ determining the mass and R-charges of the black hole
\bse\label{5d-ADM-mass-charge}%
\begin{align}
\Delta & = M\,L =  \frac{\pi L}{4 G_N^{(5)}} \left(\frac{3}{2} \mu + q_1 + q_2 + q_3 \right), \\
J_i &=\frac{\pi L}{4 G_N^{(5)}} \tilde q_i = \frac{\pi L}{4 G_N^{(5)}}\,\sqrt{q_i ( \mu+q_i)}
\end{align}
\ese%
in terms of the five dimensional Newton's constant%
\be\label{5g-GN}%
G_N^{(5)}=\frac{G^{(10)}_N}{(\pi^3 L^5)}=\frac{\pi}{2}\ \frac{L^3}{N^2}.%
\ee%
The $\mu=0$ case corresponds to $\tilde{q}_i=q_i$ and $\Delta = J_1+J_2+J_3$. This is the BPS limit. Thus $\mu$ measures the deviation from BPSness. These singular configurations were interpreted as distributions of smeared giant gravitons in \cite{Myers:2001aq}, where the flux quantisation conditions
\begin{equation}\label{Ni-Ji}%
\frac{N_i}{N}=\frac{2J_i}{N^2}=\frac{\tilde q_i}{L^2}, \quad \quad i=1,2,3%
\end{equation}%
were derived for each of the three types of giants supporting these black holes. Here $N_i$ is the number of giant gravitons in each stack. Since each giant type involves a different 3-cycle in the transverse 5-sphere \cite{giants}, pairs of giants belonging to different types intersect on circles. This observation was used in \cite{Balasubramanian:2007bs,Fareghbal:2008ar,Gubser:2004xx} to argue that two R-charged AdS${}_5$ black holes should allow a dual 2d CFT description defined on the S${}^1$ where giants intersect and with central charge proportional to the total number of such intersections, i.e. $c\sim N_iN_j$ $i\neq j$. This interpretation will play an important role when we discuss the near BPS regime.

\paragraph{Extremality vs charges:}  For completeness, we review the conditions under which {\it finite} extremal R-charged black holes appear \cite{Cvetic:1999xp}:
\begin{itemize}
\item[a)] For {\it single} R-charge configurations characterised by $\{\mu,\, q_1\}$,
the condition for extremality coincides with the condition for the black hole to
be BPS, \ie $\mu\to 0$. However, as one may easily check, no local AdS${}_3$ geometry appears
as one takes the near-horizon limit. The situation is similar to Myers-Perry black
holes with two or more of the angular momenta vanishing (\emph{cf}. the discussions in the Appendix).
\item[b)] For {\it two} R-charge configurations characterised by $\{\mu,\, q_2,\,q_3\}$, horizons exist
for $\mu>\mu_c\equiv q_2q_3/L^2$. Extremality is achieved when $\mu=\mu_c$. Thus, the scale $\mu-\mu_c$ measures the amount of non-extremality.
\item[c)] For {\it three} R-charge black hole, with three generic charges of the same order of magnitude, horizons exist for $\mu$ above a certain quantity and below which we have a naked singularity \cite{Cvetic:1999xp}. For our purposes, the is important about black holes in this class is that as soon as the extremal limit is achieved, the horizon size is necessarily {\it finite}. Thus, in this regime, EVH black holes can not appear.\footnote{If one of the charges is parametrically smaller than the other two, the three-charge system can under favorable circumstances
be viewed as a perturbation of the EVH configuration identified in b), as we will describe in detail in the following.}
\end{itemize}

\paragraph{EVH vs thermodynamics:} the energy, entropy and temperature for these black holes are \cite{Cvetic:1999xp}
\be\label{S,T-2charge-BH}%
\begin{split}%
\Delta &= L\,M = \frac{N^2}{2L^2}\,\left(\frac{3}{2}\mu + q_1 +q_2+q_3\right), \\
S &=\frac{\pi N^2}{L^3}\sqrt{(r^2_++q_1)(r^2_++q_2)(r^2_++q_3)}\ ,\\
T &=\frac{f'(r_+) r_+^3}{4\pi \sqrt{(r^2_++q_1)(r^2_++q_2)(r^2_++q_3)}}\,,
\end{split}
\end{equation}%
where $r_+^2$ is the (outer) horizon radius, defined as the largest root of $f(r)$ in \eqref{Hif-10d}.

We are interested in studying the regime of parameters where the thermodynamical conditions \eqref{def:evh} hold. If $N$ and $L$ are fixed, the vanishing of the area requires $\prod_i(r_+^2 + q_i)\sim \lambda\to 0$. Achieving this while keeping the ratio $A/T$ finite, requires $f'(r_+) r_+^3\sim \lambda$, which includes the standard extremality condition. To study this limit carefully, we express two of the independent parameters of the solution, such as $\mu$ and $q_1$, in terms of the remaining $q_2$, $q_3$ and the outer and inner horizons $r_\pm$
\begin{equation}
\begin{split}
q_1 &= r_+^2r_-^2 C(r_\pm,q_2,q_3)\,, \\
\mu-\mu_c &=(r_+^2+r_-^2)\mu_c\,C(r_\pm,q_2,q_3)+ r_+^2r_-^2\left((q_2+q_3)C(r_\pm,q_2,q_3)-1\right)
\end{split}
\end{equation}
where the scalar function $C(r_\pm,q_2,q_3)$ is
\begin{equation}%
C(r_\pm,q_2,q_3)\equiv
\frac{L^2+q_2+q_3+r_+^2+r_-^2}{\mu_c\,L^2-r_+^2r_-^2}.
\label{eq:C}
\end{equation}%
The function $f(r)$ characterising the existence of horizons becomes
\begin{equation}%
f(r)=\frac{(r^2-r_+^2)(r^2-r_-^2)}{L^2 r^4}
(r^2 +\mu_c\,L^2\,C(r_\pm,q_2,q_3)).%
\label{eq:fr}
\end{equation}%
Whenever the charges $q_2$ and $q_3$ are parametrically larger than $q_1$, it follows that \eqref{def:evh} requires
$q_2,\,q_3\gg r^2_+\sim \epsilon^2$ with $\epsilon\to 0$ and $q_1\sim\epsilon^\alpha$ with $\alpha\geq 2$. In this regime, entropy and temperature behave like
\begin{equation} \label{jj1}
  S \sim N^2 \sqrt{\frac{\mu_c}{L^2}}\,{\epsilon}, \quad \quad
  T\,L\sim \frac{\epsilon}{\sqrt{\mu_c/L^2}} \quad \Longrightarrow \quad \frac{S}{T\,L} \sim N^2\,\frac{\mu_c}{L^2},
\end{equation}
in agreement with the general set-up described in section \ref{sec:gphil}.

There are two physically distinct situations compatible with the above regime:
\begin{itemize}
\item[1.] Near-extremal {\it near-BPS} case in which $q_2,q_3\sim \epsilon$. This corresponds to a dilute giant graviton approximation in which the black hole temperature remains finite, whereas the ratio
\begin{equation}
  \frac{S}{T\,L}\propto N_2N_3
\end{equation}
is proportional to the total number of giant graviton intersections, using the quantisation conditions \eqref{Ni-Ji}.
\item[2.] Near-extremal {\it non-BPS} case in which $q_2,q_3$ remain finite and the temperature scales to zero.
\end{itemize}

As argued in section \ref{sec:gphil}, any attempt to make the entropy finite will require to take $N\to \infty$.
In the {non-BPS case}, see section~\ref{sec:nonbps}, this will be achieved by requiring $N^2 \epsilon=$ finite but large. As we discuss in section \ref{sect:nearbps}, the {near-BPS} limit will turn out to be more subtle. Even though finite entropy is thermodynamically achieved by $N\epsilon\sim 1$, near horizon considerations and reliability of classical supergravity will instead suggest to consider $N\epsilon^2$ to be finite but large.

In the next sections, we will discuss how these physical differences are encoded in the properties of the candidate emergent 2d IR CFTs, i.e. their central charges and energy levels, after explicitly identifying the near horizon geometries of the R-charged AdS${}_5$ black holes \eqref{10-dim-general} in the two regimes described above.

\section{Near-BPS R-charged AdS${}_5$ EVH black holes}
\label{sect:nearbps}

In this section, we study the near-extremal near BPS limit
\begin{equation*}
  \mu\to 0 \,,\, \qquad \mu-\mu_c\to 0 \quad \quad {\rm with} \quad \quad L \,\,{\rm fixed}.
\end{equation*}
This forces the R-charges $q_i$ to scale to zero. Thus, we will be working in some dilute giant graviton approximation.
Following \cite{Balasubramanian:2007bs,Fareghbal:2008ar}, we consider
the near horizon limit\footnote{We have assumed $\mu_1\sim\mu_1^0\neq 1$. For a detailed discussion regarding this possibility, see \cite{Balasubramanian:2007bs,Fareghbal:2008ar}.}
\begin{equation}\label{eq:2chargelim}
\begin{split}
r = \epsilon \tilde\rho, &\qquad \theta_i=\theta_i^0-\epsilon^{1/2}
\hat \theta_i, \,\,\, 0\leq\theta_i^0\leq \pi/2, \\
   \mu-\mu_c= \epsilon^2 \hat M , &\qquad q_i=\epsilon \hat
q_i, \,\,q_1 = \epsilon^2\hat q_1\,, \qquad
\psi_i=\frac{1}{\epsilon^{1/2}}\left(\phi_i-\frac{t}{L}\right),\
i=2,3,
\end{split}
\end{equation}
while keeping $\tilde\rho,\ \hat q_i,\ \hat q_1,\hat M,\ \theta_i^0,\ L,\ \psi_i$ fixed. Choosing $\mu_1^0=\cos\theta_1^0$, the resulting near horizon metric is \cite{Fareghbal:2008ar}
\begin{equation}
  ds^2 = \epsilon\left[\mu^0_1\,ds_6^2+\frac{L^2}{R^2_S\mu_1^0} ds^2_{\CM_4}\right] \quad \quad {\rm with} \quad \quad
  R_S^4 = \hat q_2\hat q_3\equiv\hat\mu_c L^2\,.
 \label{eq:nhnearbpsmetric}
\end{equation}
Here, $ds^2_{\CM_4}$ stands for the metric
\begin{equation}
ds^2_{\CM_4} = \sum_{i=2,3}\ \hat q_i\left(d\hat\mu_i^2+(\mu_i^0)^2 d\psi_i^2\right),
\end{equation}
where $d\mu_i=-\epsilon^{1/2}d\hat\mu_i$ $i=2,3$.\footnote{In terms of an explicit parameterisation consistent with the choice $\mu_1^0=\cos\theta_1^0$, we have $d\hat\mu_2 = \cos\theta_1^0\cos\theta_2^0d\hat\theta_1-\sin\theta_1^0\sin\theta_1^0d\hat\theta_2$ and $d\hat\mu_2 = \cos\theta_1^0\sin\theta_2^0d\hat\theta_1+\sin\theta_1^0\cos\theta_1^0d\hat\theta_2$.} The 6d Lorentzian metric describes a local AdS${}_3\times {\rm S}^3$
\begin{eqnarray}\label{NH-BPS-6d}
  ds_6^2 &=& -\frac{(y^2-y_+^2)(y^2-y_-^2)}{y^2R_S^2}\,dt^2
+ \frac{R_S^2y^2}{(y^2-y_+^2)(y^2-y_-^2)}\,dy^2 +
y^2\left(d\phi_1-\frac{y_+y_-}{y^2\,R_S}dt\right)^2 \nonumber \\
& & +R_S^2 d\Omega_3^2,
\end{eqnarray}
in terms of the new radial coordinate
\begin{equation}
\label{eq:def1}
  y^2=\frac{L^2}{R_S^2}\left(\tilde\rho^2+\hat q_1\right).
\end{equation}

We want to stress that  $t$ and $\phi_1$ were {\it not} rescaled in \eqref{eq:2chargelim}, which is consistent with the finiteness of the temperature discussed in section \ref{sect2} in the near-BPS regime, and $\phi_i$ were forced to rotate at the speed of light, in $L$ units, matching the rotating velocity of the constituent giant gravitons. The coordinates $\psi_i$ parameterising deviations from this co-rotation become effectively non-compact. The near horizon geometry is characterised by two parameters:
$\hat{M}$ and $\hat{q}_1$. These describe near-extremality. It is convenient to define $\mu=\epsilon^2\hat\mu$ and $\mu_c=\epsilon^2\hat\mu_c$ so that $\hat\mu_c=\hat q_2\hat q_3/L^2$ to ease the notation below. Lastly, notice that to achieve an overall scaling in the metric, one is forced to geometrically focus on a strip $\theta\sim\theta_i^0$ of the transverse 5-sphere. We will return to this point when interpreting the entropy of the metric \eqref{eq:nhnearbpsmetric}.

Notice both $AdS_3$ and $S^3$ spaces have equal radii $\ell_3=R_S$. The parameters $y_\pm$ are determined in terms of the original black hole parameters by
\begin{equation}
\label{jj5}
\begin{split}
  y_+^2+y_-^2 &= \frac{L^2}{R_S^2}\,\left(\hat M + 2\hat q_1\right)=R^2_S\left(\frac{\hat\mu+2\hat q_1}{\hat\mu_c}-1\right)\,, \\
   y_+\,y_-& = \frac{L^2}{R_S^2}\,\sqrt{\left(\hat M + \hat \mu_c + \hat q_1\right)\,\hat q_1}=R^2_S\sqrt{\frac{\left(\hat  \mu+ \hat q_1\right)\,\hat q_1}{\hat\mu_c^2}}\,.
\end{split}
\end{equation}

Depending on the values of $\hat\mu$ and $\hat q_1$ the locally AdS$_3$ part of geometry \eqref{NH-BPS-6d} corresponds to different quotients of  AdS$_3$ (for example see \cite{AdS3-CFT2} or Appendix B of \cite{Fareghbal:2008ar}). Note that $\hat\mu\geq 0$ and $\hat q_1\geq 0$ in our conventions, which implies that $\hat M/\hat\mu_c\geq -1$ ($\hat M$ can be negative). \begin{itemize}
\item
For the special case of $\hat\mu=0\ (\hat M/\hat\mu_c=-1),\ \hat q_1=0$ we have global AdS$_3$. This corresponds to the near horizon limit of 1/4 BPS AdS$_5$ black hole.
\item When $\hat M\leq -2(\sqrt{\hat q_1(\hat\mu+\hat q_1)}+\hat q_1),\ \hat M^2\geq 4\hat\mu_c\hat q_1$ we have a conic space. This can happen when $\hat M+4\hat q_1\leq 0$ and $4\hat q_1\leq\hat \mu_c$. The 1/8 BPS AdS$_5$ black hole with $\mu=0, 4q_1\leq \mu_c$ falls in this class.
\item
For  any real value of $y_\pm$ corresponding to $\hat M^2\geq 4\hat q_1\hat\mu_c,\ \hat M\geq 0$ we have BTZ black holes.  For the special case of $\hat q_1=0,\ \hat\mu=\hat\mu_c$ we have massless BTZ and for $\hat q_1=\hat M^2/(4\hat\mu_c)$ we have extremal BTZ.  The  mass  and angular momentum of the BTZ black hole is given by
\begin{equation}\label{BTZ-MJ-BPS}
  M_{\textrm{BTZ}} = \frac{y_+^2+y_-^2}{8G_N^{(3)}\,\ell_3^2}, \quad \quad J_{\textrm{BTZ}} = \frac{2y_+y_-}{8G_N^{(3)}\,\ell_3}\,.
\end{equation}
\item Finally for $\hat M^2< 4\hat q_1\hat\mu_c$ and $|\hat M+2\hat q_1|< 2\sqrt{\hat q_1(\hat \mu+\hat q_1)}$ the geometry has a naked singularity. The special case of 1/8 BPS R-charged black hole  $\mu=0,\ q_1>4\mu_c$ falls into this class. This geometry can correspond to BPS rotating D-string like excitations in the AdS$_3$. This latter case, however, should be explored in more detail, which we postpone to future works.

\end{itemize}

\paragraph{Black hole vs near horizon entropies:} Using \eqref{S,T-2charge-BH}, the entropy of the full black hole in the limit of charges defined in \eqref{eq:2chargelim} equals
\begin{equation}\label{entropy-NH-BPS}%
S =\pi\frac{R^3_S}{L^4}\,y_+ (N\epsilon)^2,
\end{equation}
where we already used \eqref{eq:def1}. It is not surprising to check that the entropy of the near horizon geometry \eqref{eq:nhnearbpsmetric} does {\it not} match this result
\begin{equation}
  S_{{\rm 3d}} = \frac{2\pi y_+}{4G_3} = 8S\,\mu_2^0\mu_3^0\,{\cal V}_{\CM_4},
\label{eq:3dbpsent}
\end{equation}
where $\V4$, which is kept finite in the limit, is defined in \eqref{V4}. This mismatch is physically expected because in taking the near horizon limit \eqref{eq:2chargelim} we were forced to focus on a strip in the transverse 5-sphere.
To derive this result, some comments and definitions are in order:
\begin{itemize}
\item[1.] By construction, the volume of the flat non-compact manifold $\CM_4$ is infinite. Since the local coordinates describing the latter are $\epsilon$ dependent, one can provide a natural regularisation by keeping $\epsilon$ very small, but finite. This gives rise to
\begin{equation}\label{V4}
  {\rm vol}\CM_4 = (2\pi)^2\,\frac{R_S^4}{\epsilon^{2}}\,\mu_2^0\mu_3^0\,\V4,
\end{equation}
which defines $\V4$.
\item[2.] The 3d Newton's constant is computed in the standard fashion, using the regularisation mentioned above, and properly dealing with the factor $\epsilon\mu_1^0$ in front of the 3d metric when comparing
\begin{equation*}
  \frac{1}{G_{10}}\int d^{10}x \sqrt{-{\rm det} g_{10}}\, {\cal R}_{10} = \frac{1}{G_3}\int d^3x \sqrt{-{\rm det} g_{3}}\,{\cal R}_{3}.
\end{equation*}
Proceeding in this way, one finds
\begin{equation}
  \frac{1}{G_3} = 16(N\epsilon)^2\,\frac{R_S^3}{L^4}\,\mu_2^0\mu_3^0\,\V4= \frac{16 N^2\epsilon^4}{\ell_3}\ \frac{{\rm vol}\CM_4}{(2\pi)^2 L^4}\,,
\label{eq:3dGN}
\end{equation}
justifying \eqref{eq:3dbpsent}.
\end{itemize}

It is interesting to emphasise that focusing on a `strip' of a black hole horizon, when taking the near horizon limit, is generic in non-extremal black holes. The difference is that the 2d geometry close to the generic non-extremal horizon is Rindler, whereas the (near)-BPS and (near) EVH case studied here give rise to AdS${}_3$ (BTZ). Technically, this occurs to guarantee analyticity in the $\epsilon$ expansion of the black hole metric components when taking its near horizon limit. The latter ensures the limiting metric remains a solution to supergravity equations. Conceptually, if one thinks of the horizon as the location where the black hole degrees of freedom live (at least from the perspective of an observer at infinity), it is clear that such near horizon description will never reproduce the correct entropy because the latter loses the information on the curvature of the original horizon by approximating it with a flat tangent plane.

Gravitationally, and for the reasons just mentioned, it is natural to interpret the so obtained near horizon Bekenstein-Hawking entropy \eqref{eq:nhnearbpsmetric} as an {\it entropy density}. This is more even so in the particular example discussed in this subsection, given the microscopic interpretation of the BPS R-charged black holes as a distribution of smeared giant gravitons on the 5-sphere \cite{Myers:2001aq} and the arguments provided in \cite{Balasubramanian:2007bs} identifying the open strings stretched between these giants as responsible for the entropy of their near-BPS limits. The dependence on the point where the strip lies, i.e. $\mu_2^0\mu_3^0$, provides the natural measure where to integrate such density. Not surprisingly, one finds
\begin{equation}\label{theta-integral}%
\int_{\mu_2^2+\mu_3^2\leq 1} \mu_2\mu_3 d\mu_2 d\mu_3=\frac{1}{8}.
\end{equation}
That is, if we suitably sum over the entropies of each BTZ black hole located at different strips (different values of $\mu_2, \mu_3$), given in \eqref{eq:3dbpsent}, we recover the entropy of the original 5d black hole.
Unfortunately, it is not clear to us what the process of integrating over this entropy density means in the language of 2d CFTs that naturally would arise as the dual descriptions of the near horizon geometries
\eqref{eq:nhnearbpsmetric}.\footnote{It is possible that requiring to have a consistent on-shell near horizon geometry  is responsible for the focusing on a horizon 'strip'. Recently, there have been discussions trying to argue that the low energy physics in (non-)extremal black holes is described by a 2d CFT, without appealing to its near horizon geometry, but to the wave equations satisfied by probe fields on the geometry \cite{More-Andy,Cvetic:2011hp}. If one would take a similar attitude in these black holes, one can envision keeping the information about the full black hole geometry. We will come back to this point in section \ref{sec:kerr-cft}, when comparing our results with the Kerr/CFT predictions.}

\subsection{Non-trivial IR dynamics and  scaling of $N$}

The entropy of the original black hole \eqref{entropy-NH-BPS} goes to zero, for finite $N$, as a consequence
of the dilute giant graviton  approximation. Given the overall $\epsilon$ scaling in the near horizon
metric \eqref{eq:nhnearbpsmetric}, it is natural to interpret the latter as a rescaling of the 10d Planck
scale, i.e. $\ell_p^4\to \epsilon^2\ell_p^4$, as we usually do in the decoupling limits leading to the AdS/CFT
correspondence \cite{ads-cft}. Keeping $L$ finite requires $N$ to scale as
$N\epsilon^2\sim 1$ if $g_s$ remains fixed, i.e. $\alpha^\prime\to\epsilon\alpha^\prime$. (Note that by $N\epsilon^2\sim 1$ we mean $N\epsilon^2$ is kept finite but large in the near horizon limit.)
This is the same scaling considered in \cite{Fareghbal:2008ar}. Given the non-compactness of the transverse space in the limit \eqref{eq:2chargelim}, it is the { entropy density} that one should require to keep { finite}
\begin{equation}
  s = \frac{S_{{\rm 3d}}}{{\rm vol} \CM_4} \propto \left(N\epsilon^2\right)^2 \,\,{\rm finite\ but\ large}  \quad \Longrightarrow \quad N\epsilon^2\sim 1.
\end{equation}
Thus, both considerations are consistent with the same scaling. We provide two further physical arguments for why $N\epsilon^2\sim 1$ can be a meaningful limit to study:
\begin{itemize}
\item[1.] One can estimate the {\it mass density} of open strings stretched between intersecting giants as%
\begin{equation}\label{open-mass}%
m_{{\rm open}}= \frac{M_{{\rm open}}}{{\rm length}\,\CM_4} \sim \epsilon^{1/2}\frac{L\delta\hat\theta}{l_s^2}\frac{1}{L\epsilon^{-1/2}}\sim \frac{\sqrt{g_s}}{L^2} \delta\hat\theta \sqrt{N\epsilon^2}\,.%
\end{equation}
where we used $L^4=4\pi g_s l_s^4N$ and kept $g_s$ fixed. Thus, requiring energy density finiteness of these excitations also dictates the scaling $N\epsilon^2\sim 1$.
\item[2.] The smallest distance computed in the near horizon metric \eqref{eq:nhnearbpsmetric} is of order $\ell\sim L\,\sqrt{\epsilon}$ and the curvature invariants of the near horizon metric \eqref{eq:nhnearbpsmetric} are of order $\ell^{-2}$. In order to have a valid supergravity approximation in which stringy corrections are small we need to require $\ell\gtrsim l_s$. Since $\ell/\ell_s\sim \left(N\epsilon^2\right)^{1/4}$, for a fixed $g_s$, validity of supergravity leads to $N\epsilon^2\sim 1$.
The validity of the supergravity description also demands  $sL^4\gtrsim 1$ where $s$ is the entropy density. This latter, as discussed above, is also satisfied with $N\epsilon^2\sim1 $ scaling.\footnote{One could have considered scaling $N\epsilon\sim 1$ so that the entropy, and not its density, remains finite. However, as pointed out, with such a scaling the overall scale of the near horizon metric will be $L/\sqrt{N}$ which is much smaller than string scale
and the gravitational description is no longer valid.}

\end{itemize}

\subsection{Comments on  ${\cal N}$=4 SYM and 2d CFT descriptions}

Our near BPS black holes have a UV description as thermal states in ${\cal N}$=4 SYM. On the other hand, our AdS${}_3$ near horizon geometry hints at the emergence of an IR 2d CFT dual description. Here we advocate a more direct connection between the two CFTs and will make some further comments in section \ref{sec:nearbpskerr}.

The evaluation of the gravitational charges $\{M,\,J_i\}$ in the near BPS limit \eqref{eq:2chargelim} identifies the sector of the Hilbert space in ${\cal N}=4$ SYM that we are focusing on:
\begin{subequations}\label{Delta-Ji-BPS}
\begin{align}%
\Delta &=\frac{N^2\epsilon}{2L^2}\left(\hat q_2+\hat q_3 +
\frac{\epsilon}{2}\left({3\hat\mu_c+2\hat q_1+3\hat M}\right)\right), \\
J_1 &= \frac{(N\epsilon)^2}{2L^2}\,\sqrt{\hat q_1\,\left(\hat\mu_c +\hat M + \hat q_1\right)}, \\
J_i &= \frac{N^2\epsilon}{2L^2}\,\left(\hat q_i +
\frac{\epsilon}{2}\left(\hat\mu_c + \hat M\right) +
\CO(\epsilon^2)\right), \quad \quad i=2,3\,.
\end{align}
\end{subequations}
In the regime $N\epsilon^2\sim 1$ where the entropy density remains finite, two remarks are in order:
\begin{itemize}
\item[1.] The dominant divergent contributions to energy $\Delta$ and R-charges $J_i$ scale like $N^{3/2}$ \cite{Fareghbal:2008ar}.
\item[2.] All quantities measuring the magnitude of the deviation from the BPS EVH solution, $J_1$ and $\Delta-J_2-J_3$, diverge like $\left(N\epsilon\right)^2$, while their {densities} $J_1/{\rm vol}\,\CM_4$ and $(\Delta-J_2-J_3)/{\rm vol}\,\CM_4$ remain {finite}. The latter are related to the mass and
angular momentum of the BTZ black holes geometries obtained in the near horizon limit.
\end{itemize}

Here, we propose an interpretation along the lines of BMN \cite{BMN}. In that case, the geometrical limit is a Penrose limit \cite{Penrose}
corresponding to focusing the dynamics onto a sector of ${\cal N}=4$ SYM where $\Delta$ and $J$ scaled like $N^{1/2}$ keeping $\Delta-J$ finite. Similarly, our limit \eqref{eq:2chargelim} identifies the set of ``almost-quarter-BPS'' operators characterised by \cite{Fareghbal:2008ar}
\begin{equation}\label{BMN-BPS}%
\begin{split}
\Delta, J_2, J_3 \sim N^{3/2}\,& ,\qquad \lambda_{\textrm{'t Hooft}}=g^2_{YM} N\sim N\to\infty\\
\Delta-J_2-J_3,\,J_1\sim N\,&, \quad \lambda_{{\rm eff}}= \frac{g^2_{YM}N}{N_2N_3}\,,
\end{split}%
\end{equation}%
where $N_2, N_3\sim N^{1/2}$ are the numbers of giants.
As explained earlier, these degrees of freedom are expected to be associated with open strings
stretched between smeared giant gravitons rather than with closed strings in the bulk \cite{Balasubramanian:2007bs}.
It is not clear to us whether simplifications similar to those that appear in the standard BMN set-up will occur here, but it is interesting to point out that preliminary steps in this direction have been taken in \cite{deMello}.

If we take the appearance of AdS$_3$ and BTZ geometries as serious evidence of the existence of a dual
IR CFT, we can use
the standard AdS3/CFT2 dictionary to connect
the BTZ mass and angular momentum to $L_0$ and $\bar L_0$ of this conjectured dual CFT \cite{AdS3-CFT2}
\be\label{L0-barL0}%
\begin{split}
L_0-\frac{c}{24}&=\frac{M_{BTZ}\ell_3+J_{BTZ}}{2}=\frac{c}{24}\left(\frac{y_++y_-}{\ell_3}\right)^2\,,\\
\bar L_0-\frac{c}{24}&=\frac{M_{BTZ}\ell_3-J_{BTZ}}{2}=\frac{c}{24}\left(\frac{y_+-y_-}{\ell_3}\right)^2\,,
\end{split}\ee%
where we have used \eqref{BTZ-MJ-BPS}. It is straightforward to check that \eqref{Delta-Ji-BPS} yields
\be
{\Delta-J_2-J_3\pm J_1}= \frac{N^2\epsilon^2}{4}\left(\frac{\ell_3}{L}\right)^4\ \left(\frac{y_+\pm y_-}{\ell_3}\right)^2\,.
\ee
We may then identify
\be
\frac{L^4}{{\rm vol}\CM_4}(\Delta-J_2-J_3+ J_1)=L_0-\frac{c}{24}\,,\qquad \frac{L^4}{{\rm vol}\CM_4}(\Delta-J_2-J_3-J_1)=\bar L_0-\frac{c}{24}\,,
\ee
and
\be
c=6N_2N_3\frac{L^4}{{\rm vol}\CM_4}\,,
\ee
where $N_2, N_3$ are the numbers of giants. The central charge $c$ has essentially the same form as in usual D1-D5 system, but now the central charge is proportional to density of giant gravitons;
$L_0\pm \bar L_0$ are examples of the quantities $\Delta_{IR}$ and $J_{IR}$ discussed in section \ref{sec:gphil}. We also note that for all vales of $\hat\mu, \hat q_1\geq 0$, $L_0, \bar L_0$ have a non-negative spectrum and hence the proposed 2d CFT is unitary. The vacuum of the 2d CFT, $L_0=\bar L_0=0$ corresponds to the 1/4 BPS black hole with $\mu=q_1=0$.

\section{Non-BPS R-charged AdS${}_5$ near-EVH black holes}
\label{sec:nonbps}

In this section, we analyze the non-BPS version of the near EVH limit described in section \ref{sect:nearbps}. This requires studying $\mu-\mu_c\to 0$ keeping $\mu_c$ finite. Thus, the charges $q_i$ $i=2,3$ will be kept finite. Since the deep interior of these non-BPS extremal black holes resembles the one of massless BTZ black holes, the discussion in \cite{EVH-BTZ} will apply. Thus, there will exist two different near horizon limits: one giving rise to a pinching AdS${}_3$ orbifold, where the periodicity of the compact dimension goes to zero and a second giving rise to the null self-dual orbifold \cite{c-h}. We want to stress that a priori, one may have not expected any decoupled geometry in this regime given the unbounded nature of the Hamiltonian in this sector. We leave this point to the discussion section.

\paragraph{Emergence of pinching orbifolds:} Since pinching orbifolds allow to explore the physics near extremality,\footnote{This was shown in some detail in the appendix of \cite{EVH-BTZ}.} we will parameterize the outer $(r_+)$ and inner $(r_-)$ horizons in \eqref{eq:fr} in terms of
\begin{equation}
r_{\pm}^2 = r_{\ast}^2 \pm \delta r_{\ast}^2,%
\end{equation}
in the limit $r_\ast,\delta r_\ast\to 0$. We define the non-BPS near-EVH near-horizon limit as
\begin{align} \label{jj21}
  r=\epsilon\,\frac{(q_2q_3)^{1/4}}{L}\,y\,, \quad r_\ast &=\epsilon\,\frac{(q_2q_3)^{1/4}}{L}\,\rho_\ast\,, \quad \delta r_\ast = \epsilon\,\frac{(q_2q_3)^{1/4}}{L}\,\delta\rho_\ast\,, \nonumber \\
  t=-\frac{L}{(q_2q_3)^{1/4}}\,\frac{\tau}{\epsilon}\,, \quad \phi_1 &=
  \frac{\varphi}{\epsilon}\,, \quad \phi_i = \psi_i - \frac{\tilde{q}_i}{q_i}\frac{1}{(q_2q_3)^{1/4}}\frac{\tau}{\epsilon}\,,\,\,\,i=2,3 \quad \nonumber \\
 q_1 = \epsilon^4\,\hat q_1\,, \quad \mu - \mu_c &= \epsilon^2\,\hat M\,, \quad \quad \epsilon\to 0,
\end{align}
keeping all parameters and coordinates in the right hand sides fixed. The resulting metric is \cite{Fareghbal:2008ar}
\begin{equation}
  ds^2 = \mu_1\,ds^2(\ell_3) + \frac{L^2}{\mu_1}\,ds^2_{{\cal M}_7}
\label{eq:nonbpsmetric}
\end{equation}
This involves a 7d Euclidean compact metric
\begin{equation}\label{M7-metric}
  ds^2_{{\cal M}_7} = \frac{\sqrt{q_2q_3}}{L^2}\,\mu_1^2\,d\Omega_3^2 + \sqrt{q_2/q_3}\,(d\mu_2^2 + \mu_2^2\,d\psi_2^2) +  \sqrt{q_3/q_2}\,(d\mu_3^2 + \mu_3^2\,d\psi_3^2),
\end{equation}
and a 3d lorentzian locally AdS${}_3$ metric
\begin{equation}
ds^2(\ell_3)=-\frac{(y^2-y_+^2)(y^2-y_-^2)}{y^2\ell_3^2}\,d\tau^2 +
\frac{\ell_3^2y^2}{(y^2-y_+^2)(y^2-y_-^2)}\,dy^2 +
y^2\left(d\varphi+\frac{y_+y_-}{y^2\,\ell_3}d\tau\right)^2,
\label{eq:lbtz}
\end{equation}
having radius $\ell_3$
\be\label{r02}%
 \ell_3^2 = \sqrt{q_2q_3}\,\frac{L^2}{h^2},\qquad h^2\equiv L^2 + q_2 + q_3, %
\ee%
and a periodic $\varphi$ satisfying $\varphi \sim \varphi + 2\pi\epsilon$.

The near-horizon geometry depends on two parameters $\hat{M}$ and $\hat{q}_1$. Depending on their values, the 3d geometry describes an extremal, non-extremal, massive, or massless BTZ geometry with outer and inner horizons $y^2_\pm=\rho_\ast^2 \pm \delta\rho_\ast^2$ given by\footnote{The geometry is BTZ black hole if $(y_+\pm y_-)^2\geq 0$. If $(y_+\pm y_-)^2<0$ we have a conic space and if $(y_++ y_-)^2>0, \ (y_+-y_-)^2<0$ we have a time-like naked singularity. The latter two are only possible when the original R-charged black hole violates the extremality bound.}
\begin{equation}\label{y-pm}
\left(y_+\pm y_-\right)^2 = \frac{L^4}{h^2 \sqrt{q_2q_3}}\left(
{\hat M}\pm 2\frac{h}{L^2}\sqrt{{\hat q_1}q_2q_3}\right)\,.
\end{equation}

It was shown in \cite{Fareghbal:2008ar} that there exists a non-trivial consistent truncation of type IIB with a constant dilaton, metric and self-dual 5-form to six dimensions of the form
\begin{equation}
  ds^2 = \mu_1 ds^2_6 + \frac{L^2}{\mu_1}\left(\sqrt{q_2/q_3}\,(d\mu_2^2 + \mu_2^2\,d\psi_2^2) +  \sqrt{q_3/q_2}\,(d\mu_3^2 + \mu_3^2\,d\psi_3^2)\right)\ .
\end{equation}
Notice this is indeed of the form found above with $ds_6^2 = ds^2(\ell_3) + \sqrt{q_2q_3}/L^2 d\Omega_3^2$, where the 6d part is a space of negative constant scalar curvature $R_6=-\frac{6(q_2+q_3)}{L^2\sqrt{q_2q_3}}$.

The linear $\epsilon$ periodicity in $\varphi$ affects the ADM mass and angular momentum of the pinching BTZ geometries described above \cite{BTZ}
\begin{equation}\label{jj23}%
M_{\text{BTZ}} = \frac{y^2_++y^2_-}{8\ell_3^2
G_N^{(3)}}= \frac{N^2\epsilon}{4}\frac{\hat M}{L^2(q_2q_3)^{1/4}}\,,
\qquad   J_{\text{BTZ}} =\frac{y_+y_-}{4\ell_3 G_N^{(3)}}\epsilon=
\frac{N^2\epsilon}{2}\,\sqrt{\frac{\hat q_1{q_2q_3}}{L^6}},%
\end{equation}%
where we used the value for the 3d Newton's constant $G_N^{(3)}$ \cite{Fareghbal:2008ar}
\begin{equation}\label{3d-GN-non-BPS}
\frac{1}{G_N^{(3)}} =
\frac{L^3}{G_N^{(10)}}\,\left(q_2q_3\right)^{3/4}\,(2\pi)^2
\left(\text{vol} S^3\right)\,\int \mu_2\mu_3d\mu_2d\mu_3=
2N^2\,\frac{\left(q_2q_3\right)^{3/4}}{L^4}.
\end{equation}
The Hawking temperature is finite and $\epsilon$ independent
\begin{equation}\label{jj10}
T_{\text{BTZ}} = \frac{(y_+^2-y_-^2)}{2\pi\ell_3^2\,y_+}= \frac{h^2\,\delta\rho_\ast^2}{\pi\,L^2\,\sqrt{q_2q_3}\,\sqrt{\rho_\ast^2+\delta\rho_\ast^2}},
\end{equation}
since it can be computed requiring the absence of conical singularities in its Euclidean continuation. Notice $T_{\text{BTZ}}$ differs from the 5d black hole temperature $T$  \eqref{S,T-2charge-BH} by
\begin{equation} \label{jj11}
T=\frac{{(q_2q_3)^{1/4}}}{L}\,T_{\text{BTZ}}\epsilon.
\end{equation}
This relation is consistent with the scaling of the time coordinate $t$ in the near horizon limit \eqref{jj21}.

\paragraph{The null orbifold appearance:} Following the discussion in \cite{EVH-BTZ}, there should exist a second inequivalent near horizon limit when we restrict ourselves to deformations preserving extremality
\begin{equation}
  y_+=y_- \quad \quad \Longleftrightarrow \quad \quad {\hat M} = 2\frac{h}{L^2}\sqrt{{\hat q_1}q_2q_3}.
\end{equation}
Indeed, for this subset of excitations, we can modify the singular large gauge transformations appearing in \eqref{jj21} to
\begin{eqnarray}
  t&=&-\frac{L}{(q_2q_3)^{1/4}}\,\frac{\tau}{\epsilon^2}, \quad \quad \phi_1 = \varphi - \frac{\tau}{\ell_3\,\epsilon^2}, \label{eq:nonbpsnull} \\
  \phi_i &=& \chi_i +\frac{\tilde q_i}{q_i^2L^2}(q_2q_3)^{1/4}y_+^2\tau - \frac{\tilde{q}_i}{q_i}\frac{1}{(q_2q_3)^{1/4}}\frac{\tau}{\epsilon^2},\,\,\,i=2,3\,. \nonumber
\end{eqnarray}
Note that in the above expression $\tilde q_i=\sqrt{q_i(\mu+q_i)}$ has an expansion in powers of $\epsilon$ because $\mu=\mu_c+\epsilon^2 \hat M$.
 The resulting metric is as in \eqref{eq:nonbpsmetric}, but there are some important differences:
\begin{itemize}
\item[a)] The 3d metric corresponds to the null selfdual orbifold
\begin{equation}\label{eq:nullselfdual-metric}
  ds^2(\ell_3) = \ell_3^2\,\frac{d\sigma^2}{4\sigma^2} + 2\sigma\,d\varphi\,\frac{d\tau}{\ell_3},
\end{equation}
where $\sigma = y^2 - y_+^2$.
\item[b)] The null AdS${}_3$ orbifold is non-trivially fibered over the 7d transverse space by replacing the $d\psi_i$  in \eqref{M7-metric} with $d\chi_i- A_i \sigma d\tau$, with $A_i=\frac{\sqrt{\mu_c q_i}}{q_i^2L^2} (q_2q_3)^{1/4}$. This turns on some constant electric fields in the transverse space.
\item[c)] The periodicity in $\varphi$ remains $\epsilon$ independent. Thus, this limit involves no pinching.
\end{itemize}

\subsection{Non-trivial IR dynamics and  scaling of $N$}
\label{sec:IRnonbps}

In this section, we want to reexamine the existence of non-trivial dynamics in the IR limits taken in \eqref{jj21} and \eqref{eq:nonbpsnull}. Since the near horizon geometries so obtained are equivalent to the ones studied for (near) extremal BTZ black holes  \cite{us,EVH-BTZ}, the viewpoint we adopt here is to assume the existence of a 2d CFT, that we shall refer to as ``parent" CFT and which will capture some of the IR dynamics. The different large gauge transformations involved in \eqref{jj21} and \eqref{eq:nonbpsnull} correspond to focusing on different sectors in the same theory. Thus, different subsectors of ${\cal N}=4$ SYM share some features with certain subsectors of these "parent" CFTs. Whenever the pinching AdS${}_3$ emerges, both chiral sectors of this CFT are decoupled \cite{EVH-BTZ}, whereas when $\varphi$ remains $2\pi$ periodic, the time coordinate scaling \eqref{eq:nonbpsnull} is appropriately interpreted as an infinite boost \cite{Balasubramanian:2003kq}, allowing us to interpret the near horizon as describing the DLCQ limit of the original non-chiral 2d CFT \cite{us,Balasubramanian:2003kq}. The null selfdual orbifold corresponds to the $p^+=0$ sector of the latter \cite{EVH-BTZ,Vijay-Simon}.

Consider the non-BPS IR limit \eqref{jj21} first. From a gravitational point of view, since \eqref{jj21} does not involve any focusing in the transverse dimensions, the entropy evaluated in the near horizon pinching geometry, as expected, equals the one of the original black hole
\begin{equation}\label{S-non-BPS}%
S=\pi\frac{\sqrt{q_2q_3}}{L^2}\,\frac{r_h}{L}\,N^2=S_{{\rm 3d}}= \frac{2\pi\epsilon\,y_+}{4G_3} =\pi \frac{(q_2q_3)^{3/4}}{L^3}\frac{y_+}{L}\,N^2\epsilon.
\end{equation}
In the original black hole geometry, the linear $\epsilon$ scaling is due to the smallness of the horizon, i.e. $r_h\sim\epsilon$.
In the 3d near horizon geometry, it is due to the pinching. Thus, the entropy vanishes in the limit $\epsilon\to 0$ while keeping $N$ fixed.

It is standard to reproduce the gravitational entropy of an AdS${}_3$ throat using Cardy's formula \cite{andy-btz}. In the presence of a non-trivial pinching, this may be a bit more subtle. When computing the mass and angular momentum of the 3d pinching geometry \eqref{jj23}, we took into account the non-trivial periodicity of the S${}^1$ circle. Using the same spacetime perspective, the standard Brown-Henneaux central charge \cite{Brown-Henneaux} will also acquire such a linear $\epsilon$ dependence
\begin{equation}\label{c-Cardy}%
c_{{\rm AdS}_3}=\frac{3\ell_3}{2G_3}\epsilon=
\frac{3\mu_c}{L\,h} N^2\epsilon.
\end{equation}
This agrees with the {\it spacetime} CFT approach described in \cite{Martinec:2001cf}, when describing AdS${}_3$ orbifolds. In this approach, both the central charge and the excitations $L_0-c/24$ and ${\bar L}_0-c/24$ scale like $N^2\epsilon$
\begin{equation}\label{L0-L0bar}%
L_0+\bar L_0-\frac{c}{12}=\frac{N^2\epsilon}{4} \frac{\hat M }{L h}\,, \qquad
L_0-\bar L_0=\frac{N^2\epsilon}{2}\sqrt{\frac{\hat q_1q_2q_3}{L^6}}\,.%
\end{equation}

Having assumed the existence of a dual 2d parent CFT, the near horizon limit \eqref{jj21} corresponds to an IR limit of such theory in which both chiral sectors are decoupled and we are left with no dynamics \cite{EVH-BTZ}. This is a direct consequence of the pinching appearing in the near horizon geometry \eqref{eq:nonbpsmetric}. This conclusion is true while holding Newton's constant fixed, but if we allow the latter to scale to zero as the near horizon (IR) limit is taken, one can keep the central charge and the entropy finite \cite{EVH-BTZ}. This is explicit in our set-up since the rank of the $\SU(N)$ gauge group in the original ${\cal N}=4$ SYM controls the 3d Newton's constant \eqref{3d-GN-non-BPS}. In other words, we can scale $N\to \infty$, keeping $L$ {\it fixed} so that $N^2\epsilon$ remains {\it finite}
\begin{equation}
  N\to \infty \quad \quad {\rm with} \quad \quad L,\,N^2\epsilon \,\,\,{\rm fixed}
\end{equation}
It is manifest that in such a double scaling limit all relevant thermodynamical and conformal field theory quantities will remain {\it finite}. (Note that since the volume of the ${\cal M}_7$ manifold remains finite, both 3d and 10d Newton constants scale to zero in the same way, as $N^{-2}$.)

We can now consider the same EVH triple scaling limit as in \cite{SheikhJabbaria:2011gc}:
\be\label{EVH-triple}
A,\ T,\ G_N \to 0\,,\qquad \frac{A}{G_N},\ \frac{A}{T}\ \ {\rm finite}\,,
\ee
where $A/G_N$ is the entropy of the EVH black hole (reproduced by Cardy formula of the dual 2d CFT) and the ratio $A/T$ (up to numerical coefficients) is equal to the central charge of the dual CFT.

The spacetime CFT perspective used in \eqref{c-Cardy} and \eqref{L0-L0bar} can now be understood as the orbifold projection of the dual parent 2d CFT \cite{Martinec:2001cf,EVH-BTZ}.  The pinching orbifold can then be understood as an {\it ensemble} at temperature $T\sim \epsilon$ describing excitations $L_0-c/24,\,{\bar L}_0-c/24\sim N^2\epsilon^2$ in this CFT theory with central charge
\begin{equation}
  c_{{\rm CFT}} = \frac{3\ell_3}{2G_3} = \frac{3\mu_c}{L\,h} N^2.
\label{eq:parentcft}
\end{equation}
We will comment on the relation between this parent CFT and the one emerging in Kerr/CFT in section \ref{sec:kerr-cft} where we also
remark on the dual 2d (parent) CFT description of the null orbifold obtained through \eqref{eq:nonbpsnull}.

\subsection{$\CN=4$ SYM interpretation}

Given the dictionary between bulk charges and ${\cal N}=4$ SYM quantum numbers, we can identify the sector of the ${\cal N}=4$ Hilbert space being explored in the near horizon limit \eqref{jj21}
\begin{subequations}\begin{align} \label{jj22}
 \Delta &= \frac{N^2}{2L^2}(\frac32\mu_c+q_2+q_3) + \frac{3\hat M}{4L^2}\,(N\epsilon)^2\,, \\
 J_i &= \frac{N^2}{2L^2}\,\sqrt{q_i\left(\mu_c+q_i\right)} + \frac{\hat M}{4L^2}\,\sqrt{\frac{q_i}{\mu_c+q_i}}\,(N\epsilon)^2\,, \quad i=2,3 \\
 J_1 &= \frac{\sqrt{\mu_c\,\hat q_1}}{2L^2}\,(N\epsilon)^2\,.
\end{align}
\end{subequations}
Any bulk probe field has a Fourier expansion in terms of both the UV and IR isometries. In the UV theory, the conformal dimension $\Delta_{{\rm UV}}$ and R-charges $J_a$ $a=1,2,3$ will naturally be related to the eigenvalues of the vector fields
\begin{equation}
  \Delta_{{\rm UV}}=iL\frac{\partial}{\partial t},\quad \quad J_a=-i\frac{\partial}{\partial \phi_a},\quad a=1,2,3.
\end{equation}
Similarly, in the IR theory, there are natural vector fields to use, related to the UV ones through the non-trivial singular large gauge transformations included in \eqref{jj21}, giving rise to the IR eigenvalues $\Delta_{{\rm IR}}$ and ${\cal J}$
\bse\label{Delta-J-extremal}%
\begin{align}%
\Delta_{{\rm IR}}&\equiv -i\ell_3\frac{\partial}{\partial
\tau}=-\frac{\ell_3}{(q_2q_3)^{1/4}\ \epsilon}
\left(\Delta-\frac{2L^2}{N^2}\sum_{i=2,3} \frac{J_i^2}{q_i}\right)\\
{\cal J}&\equiv -i\frac{\partial}{\partial
\varphi}=\frac{J_1}{\epsilon}\,.
\end{align}%
\ese%

${\cal E}$ and ${\cal J}$ may directly be related to the 2d CFT
charges.\footnote{The second equality in (\ref{Delta-J-extremal}a)
may be understood in an intuitive way. The BPS giants are spherical branes moving with speed of light on a circle in the $S^5$. Similarly  the near-extremal (far from BPS case) could be interpreted as massive (topologically) spherical 3-branes which are
 behaving like \emph{non-relativistic}
objects which are rotating with angular momentum $J_i$ over circles
with radii $R_i$, $R_i^2=\frac{L^2}{(q_2q_3)^{1/2}} q_i$
\eqref{M7-metric}. Therefore, the kinetic energy of this rotating
branes is proportional to $\sum J^2_i/q_i$.} Explicitly, one can see that
\be\label{E-J-nonBPS}%
\Delta_{{\rm IR}}={\cal E}-{\cal E}_c=L_0+\bar L_0-\frac{c}{12}\,,\qquad {\cal J}=L_0-\bar L_0\,,
\ee%
where ${\cal E}_c$ is computed for $\Delta=\Delta_c$ and
$c=c_{{\rm AdS}_3}$ is given in \eqref{c-Cardy} and  $L_0$ and $\bar L_0$
are given in \eqref{L0-L0bar}. This precise matching is intriguing,
as we are working with a sector in the $\cN=4$ SYM with large
scaling dimensions which is far from BPS and one would not expect a
protection due to supersymmetry. As another intriguing fact, although it is
not clear that the extremality bound in gravity has a specific
meaning in the $\cN=4$ SYM theory, it seems that ``extremality''
also brings some sort of protected-ness and that $\Delta=\Delta
(\mu=\mu_c),\ J_i=J_i(\mu=\mu_c)$ provides a well-defined ground
state for the BMN-type sector and possibly for a decoupled theory. This BMN-type sector, will hence contain operators with $\Delta, J_2, J_3$ of order $N^2$, while certain combination of these (given in \eqref{Delta-J-extremal}) remain finite \cite{Fareghbal:2008ar}.

Even though it would be tempting to interpret the central charge \eqref{c-Cardy} in terms of
intersecting giant gravitons, the microscopic understanding of the non-BPS regime is not established. In particular the $1/h$ factor does not have a clear origin in that framework.

\paragraph{The null orbifold case:} We can repeat the procedure for the second limit \eqref{eq:nonbpsnull} giving rise to the null orbifold.
The condition $y_+=y_-$ forces
\begin{equation*}
  \hat{M} = 2\frac{h}{L^2}\,\sqrt{\hat{q}_1\,q_2q_3}\,.
\end{equation*}
Using \eqref{jj22} and \eqref{eq:nonbpsnull} one can show that
$$
i\ell_3\frac{\partial}{\partial\tau}=\frac{N^2 q_2q_3}{4L^3h \epsilon^2}+ \frac{N^2 \sqrt{\hat q_1q_2q_3}}{L h^2}\left(\frac{q_2+q_3}{2L^2}\right)+{\cal O}(N^2\epsilon^2)\,.
$$
The first term in the above is basically what was called ${\cal E}_c$ and the term proportional to $N^2$ is measuring being out of EVH point.  One would have naively expected the $N^2$ piece not to be present. These terms may be related to the energy of electric fields discussed in  item 2. below \eqref{eq:nullselfdual-metric}. %

\section{Planar black holes as infinite charge black holes}
\label{sect5}

As discussed in section \ref{sec:gphil}, a second possibility to avoid the conclusion that extremal black holes in AdS have a dual CFT with no dynamics is to consider black holes with non-compact horizons. In the spirit of this note, we want to remind the readers of the well known result that such black holes, the so called { planar} AdS black holes, can be viewed as a large $N$ limit, i.e. an { infinite} charge limit, of global R-charged AdS black holes.

To illustrate this relation, consider an R-charged black hole in global AdS${}_{d+1}$ characterised by the radius $R$ of the $(d-1)$-sphere, its mass $M$ and and its R-charge $J$. Conformal invariance implies the equivalence between systems with parameters
\begin{equation*}
 (M,J,R) \sim (\lambda^{-1} M,J,\lambda R) \quad \quad {\rm (conformal\,\, invariance)}
\end{equation*}
Since planar AdS black holes have infinite $M$ and $J$ but their mass and charge { densities} are { finite}, one way to derive them from their global versions is to combine the rescaling
\begin{equation*}
  M\to \lambda^d M_0, \quad \quad  J\to \lambda^{d-1} J_0 \quad \quad {\rm (charge\,\, rescaling)}
\end{equation*}
with a conformal transformation in the limit $\lambda\to\infty$
\begin{equation}
  (M,J,R) \,\,\Longrightarrow\,\, (\lambda^{d-1} M_0,\lambda^{d-1} J_0, \lambda R)\quad \quad \lambda\to\infty
\end{equation}
In this way, charges go to infinity as the volume of the boundary theory, keeping their densities $(M_0,J_0)$ finite, without modifying
 the AdS curvature radius $L$.

One consequence of this procedure is to relate correlation functions of operators ${\cal O}$ in a planar AdS black
hole background to correlation functions in the original global AdS black hole
\begin{equation}
\langle {\cal O}(x) {\cal O}(y) \rangle^{\rm planar}_{M_0,Q_0} \equiv
\lim_{\lambda\rightarrow \infty} \lambda^{-2\Delta} \langle
{\cal O}(\lambda^{-1} x) {\cal O}(\lambda^{-1} y) \rangle_{\lambda^d M_0,\lambda^{d-1} Q_0}
\end{equation}
where $\Delta$ is the conformal weight of ${\cal O}$. Notice how the boundary points on the $(d-1)$-sphere where operators ${\cal O}$ are inserted get rescaled, due to conformal invariance, as one increases the mass and R-charge of the black hole. There are similar expressions for higher n-point functions.

\paragraph{Planar RN AdS black holes:} The procedure reviewed above was already applied for the R-charge AdS${}_5$ black holes \eqref{10-dim-general} in \cite{Cvetic:1999xp, planar}. Focusing on the equal R-charge black hole, i.e. $q_i=q$ $i=1,2,3$, the combined charge rescaling and conformal transformation amount to
\begin{equation}
  \mu \to \lambda^4 \mu_0, \,\,q\to \lambda^2 q_0,\,\, r\to \lambda r,\,\,t\to \lambda^{-1} t
\end{equation}
For large $\lambda$, this will rescale the ADM mass by $\lambda^4$ and the R-charge $\tilde{q}$ by $\lambda^3$, as desired. The net result of the transformation is to:
\begin{itemize}
\item[1.] replace $f(r)$ in \eqref{Hif-10d} with
\begin{equation}
f = -\frac{\mu_0}{r^2} + \frac{r^2}{L^2}\left(1+\frac{q_0}{r^2}\right)^3, %
\end{equation}
keeping $H_i$ and $\Delta$ unmodified.
\item[2.] replace the physical R-charge $\tilde{q}=\sqrt{\mu_0 q_0}$.
\item[3.] rescale the 3-sphere metric as $d\Omega_3^2 \to \lambda^2 d\Omega_3^2$, so that in the
$\lambda\to\infty$, it gets replaced by $\mathbb R^3$, making explicit the non-compactness of both the boundary theory and the black hole horizon.
\end{itemize}

The resulting metric corresponds to the planar RN AdS black hole. Its low temperature regime and its AdS${}_2\times \mathbb R^3 \times S^5$ near horizon were studied in \cite{Faulkner:2009wj} in connection with quantum criticality and the emergence of IR CFTs. Their work gives explicit evidence for the existence of non-trivial dynamics in these set-ups.

\paragraph{Identifying the AdS${}_2$ isometries in the UV theory?}

Planar black holes arise as in the large charge limit of ordinary
black holes. In addition, extremal planar black holes are presumably
unstable against backreaction as mentioned above. It would therefore
be sufficient to find approximate $SL(2)$ generators in the UV
theory. In fact, it would be sufficient to find UV generators $L_k$,
which obey
\be \label{jjj1} P_{near-extremal}(
[L_k,L_l]-(k-l)L_{k+l} )= {\cal O}(N^{-1},Q^{-1})
\ee
where $Q$
represents the charges of the extremal black hole, and the
projection on the left-hand side is a projection onto states in
Hilbert space which are very close to extremality. The near-horizon limit is then implemented by this projection operator, in the planar limit the terms of order $1/Q$ can be neglected, and in the
supergravity limit the terms of order $1/N$ can be neglected. Thus,
if (\ref{jjj1}) is satisfied, then they are candidate near-horizon isometries of planar black holes in the supergravity description. We leave a precise description of these isometries to future work.

\section{Comparison with extremal BH/CFT}
\label{sec:kerr-cft}

In this section, we investigate the behaviour of the CFT data, i.e. central charges and Frolov-Thorne temperatures, provided by the Kerr/CFT correspondence \cite{Kerr-CFT}, and its extremal black hole extension \cite{Hartman:2008pb}, when the extremal horizon size vanishes. When the latter occurs, the AdS${}_2$ near horizon responsible for the boundary conditions proposed in \cite{Kerr-CFT,Hartman:2008pb} may disappear, giving rise to the local AdS${}_3$ throats discussed in this note.

The extremal black hole/CFT correspondence \cite{Hartman:2008pb} applies to near horizon geometries of the form
\begin{eqnarray}
ds^2 &=& A \left( - \rho^2 \, dt^2 + \frac{d\rho^2}{\rho^2} \right) +
\sum_{\alpha=1}^{n-1} F_\alpha \, dy_\alpha^2 + \sum_{i,j = 1}^{n-1+\epsilon}
\tilde g_{ij}\, \tilde e_i \, \tilde e_j\,,\nn\\
\tilde e_i & = & d\phi_i + k_i \rho \, dt . \label{dgenh}
\end{eqnarray}
When certain boundary conditions are applied to these near horizon geometries, one discovers that the asymptotic symmetry group includes a single Virasoro algebra extension for each of the compact $\U(1)$ isometries $\partial_{\phi_i}$ \cite{Chow:2008dp}. Their central charges equal \cite{Chow:2008dp}
\begin{equation}
 c_i=\frac{6k_iS_{{\rm bh}}}{\pi},
 \label{eq:cccft}
\end{equation}
where $S_{{\rm bh}}$ stands for the entropy of the original {\it finite} extremal black hole. The application of Cardy's formula
\begin{equation}
  S = 2\pi\sqrt{\frac{c_i}{6}\left(L_0^i-\frac{c_i}{24}\right)} = \frac{\pi^2}{3}c_i T_i= S_{{\rm bh}},
\label{eq:chiralcft}
\end{equation}
always reproduces $S_{{\rm bh}}$ since the CFT temperature $T_i$ equals
\begin{eqnarray}
k_i=\frac{1}{2\pi T_i}\,,\qquad
T_i= - \frac{T'^0_H}{\Omega'^0_i}
\,,\quad \text{with} \quad
T_H'^0\equiv \frac{\partial T_{\textrm{H}}}{\partial r_+} \bigg| _{r_+ = r_0}
\,,\qquad
\Omega'^0_i \equiv \frac{\partial \Omega_i}{\partial r_+} \bigg| _{r_+ = r_0},
\label{eq:kerrcftdic}
\end{eqnarray}
where $T_H(r_+)$ and $\Omega_i(r_+)$ are the temperature and angular velocities on the outer horizon $r_+$ and $r_0$ stands for its extremal value.

Finite extremal R-charged AdS${}_5$ black holes fit this discussion. Since these have three $\U(1)$ isometries describing independent rotations in S${}^5$, there should exist three inequivalent chiral CFTs reproducing the black hole entropy. This system was analysed in \cite{Lu:2009gj}, where the following Frolov-Thorne temperatures were computed
\begin{equation}
  T_i = \frac{(r_0^2+q_i)^2(q_1q_2q_3+r_0^6)}{\pi\tilde{q}_i\,L\,r_0^7\,\sqrt{H_1H_2H_3(r_0)}}.
 \label{eq:FTcft}
\end{equation}
These fix the central charges through \eqref{eq:cccft}.

Because of the analysis performed in previous sections, we will assume the bulk entropy scales like $\epsilon^\gamma$ with $\epsilon\to 0$. Given the emergence of local AdS${}_3$ throats in these situations, we are interested in matching the AdS${}_3$/CFT${}_2$ dictionary to the limiting values of the Kerr/CFT predictions above and, whenever possible, interpret and justify the latter results in terms of the former. One can distinguish three different physical cases consistent with this entropy behavior and \eqref{eq:chiralcft}:
\begin{itemize}
\item[1.] {\it Finite} central charge $c_i$, but vanishing level $L_0^i-\frac{c_i}{24}\sim \epsilon^{2\gamma}$. From an AdS${}_3$ perspective, this would correspond to keeping a {\it finite} gap in the CFT, but sending the level to zero (vacuum). Notice the CFT temperature scales like the entropy $T_i\sim \epsilon^\gamma$.
\item[2.] {\it Finite} level $L_0-\frac{c_i}{24}$, but central charge scaling like $c_i\sim \epsilon^{2\gamma}$. From an AdS${}_3$ perspective, this generates an infinite gap in the CFT, keeping the level finite. Thus, the CFT temperature scales inversely to the entropy, $T_i\sim \epsilon^{-\gamma}$.
\item[3.] {\it Vanishing} central charge and level according to
\begin{equation*}
  c_i\sim \epsilon^\alpha\,, \quad \quad L_0^i-\frac{c_i}{24}\sim \epsilon^\beta \quad \quad \alpha+\beta = \gamma.
\end{equation*}
If $\alpha,\beta >0$, the system is pushed to its vacuum while generating an infinite gap. The CFT temperature scales like $T_i\sim \epsilon^{\gamma/2-\alpha}$ and remains finite when $\alpha=\beta$.
\end{itemize}

Here we identify each of the regimes described above with the different extremal vanishing horizon limits studied in previous sections.

\subsection{Near-BPS discussion}
\label{sec:nearbpskerr}

The purpose of this section is to describe the limiting behaviour of the three CFTs that reproduce the entropy for extremal finite R-charged AdS${}_5$ black holes in the near-BPS regime \eqref{eq:2chargelim} and to identify which one matches, if any, with the 2d CFT that one may associate with the near horizon AdS${}_3$ geometry \eqref{eq:nhnearbpsmetric}. Geometrically, it is expected the latter should correspond to the chiral CFT based on the Virasoro extension of the isometry direction along which giant gravitons intersect. Thus, one expects $c_1$ and $T_1$ to have different scaling behaviour from the two remaining Kerr/CFT descriptions.

Let us analyse the CFT temperatures first. In the near-BPS limit \eqref{eq:2chargelim}, these behave like
\begin{equation}
  T_1\to \frac{\sqrt{\hat{q}_2\hat{q}_3}}{\pi L}\frac{(\hat{r}_+^2 + \hat{q}_1)^{3/2}}{\hat{r}_+^4}\sqrt{\frac{\hat{q}_1}{\hat{\mu}+\hat{q}_1}}, \quad \quad T_i \sim \frac{1}{\epsilon}\,\quad i=2,3
\end{equation}
Since the black hole entropy scales like $S\sim \left(N\epsilon\right)^2$, the central charges also have two distinctive behaviors
\begin{equation}
  c_1\sim \left(N\epsilon\right)^2\,, \quad \quad c_i\sim \frac{\left(N\epsilon^2\right)^2}{\epsilon}\,\quad i=2,3
\end{equation}
For finite $N$, all central charges scale to zero, consistent with the vanishing of the bulk entropy and the dilute giant graviton approximation in which the number of physical degrees of freedom is being scaled to zero.

Consider the regime $N\epsilon^2\sim 1$ when the bulk {\it entropy density} is finite. The analogue notion in the CFT is carried by
a central charge density. Using the volume of the transverse space computed in section \ref{sect:nearbps}, $c_1/\epsilon^{-2} \sim \left(N\epsilon^2\right)^2$ remains {\it finite}, while the other two central charge densities $c_i/\epsilon^{-2} \sim \epsilon\to 0$ would still indicate a breaking down of this effective description.\footnote{Notice the $N\epsilon\sim 1$ scaling keeping the entropy finite keeps $c_1$ finite, whereas $c_i\sim\epsilon \to 0$. Thus the breaking down of these two CFTs also remains for this scaling limit.} This confirms our expectation that the surviving CFT is the one living on the intersection of giant gravitons.

Let us explore more thoroughly the Kerr/CFT description with CFT data $\left(c_1,\,T_1\right)$. To compare with the 2d CFT dual to the near horizon AdS${}_3$ geometry \eqref{eq:nhnearbpsmetric}, we must satisfy the extremality condition, i.e. $y_+=y_-$ or equivalently, $ \hat{M} = 2\sqrt{\hat{\mu}_c\hat{q}_1}$. Standard AdS${}_3$/CFT${}_2$ tells us
\begin{equation}
  T_L = \frac{y_++y_-}{2\pi\ell_3} = \frac{y_+}{\pi\ell_3}, \quad \quad T_R =  \frac{y_+-y_-}{2\pi\ell_3} = 0.
\label{eq:cfttemp}
\end{equation}
Thus, we want to compare $T_1$ with the chiral temperature $T_L$. Using \eqref{eq:def1} and \eqref{jj5}, we learn
\begin{equation*}
  y_+^2 = \frac{L^2}{\sqrt{\hat{q}_2\hat{q}_3}}\,\sqrt{\hat{q}_1(\hat{M}+\hat{\mu}_c+\hat{q}_1)} \quad \Longrightarrow \quad \hat{r}_+^2 = \sqrt{\hat{\mu}_c\hat{q}_1}.
\end{equation*}
Thus, both temperatures are equal
\begin{equation*}
  T_L = \frac{y_+}{\pi(\hat{q}_2\hat{q}_3)^{1/4}}= T_1.
\end{equation*}
Furthermore, using \eqref{eq:cccft}, the corresponding Kerr/CFT central charge equals
\begin{equation}
  c_1 = \frac{3}{\pi^2}\frac{S}{T_1} = 3N^2\,\frac{q_2}{L^2}\frac{q_3}{L^2} = 3N_2N_3,
 \label{eq:nearbpskerr}
\end{equation}
where we used the quantisation conditions \eqref{Ni-Ji}. The latter agrees with the total number of giant intersections, independently of whether $N$ is scaled or not. Notice how a proper microscopic understanding of the system quantises this specific CFT central charge reported in \cite{Lu:2009gj}.

\paragraph{Comparison with the near horizon analysis: } As stressed in section \ref{sect:nearbps}, our near horizon limit \eqref{eq:2chargelim} focused on a strip of the transverse 5-sphere. Consequently, its Bekenstein-Hawking entropy did not match the full black hole entropy \eqref{eq:3dbpsent}. This suggests that if one uses the standard dictionary between AdS${}_3$ bulk data and 2d dual CFT quantities, the central charge so obtained will only capture the number of intersecting giant gravitons present in the focused strip. Indeed, using the Brown-Henneaux central charge \cite{Brown-Henneaux} and the 3d Newton's constant \eqref{eq:3dGN}, we obtain
\begin{equation}
  \label{BH-c-BPS}
c=\bar{c}=\frac{3\ell_3}{2G_N^{(3)}}=3N_2N_3\,(8\mu_2^0\mu_3^0\,g_{{\cal M}_4}).
\end{equation}
As usual, Cardy's formula is consistent with Bekenstein-Hawking  \eqref{eq:3dbpsent},
\begin{equation}
  S_{{\rm Cardy}} = 2\pi\sqrt{\frac{c\,(L_0-c/24)}{6}} + 2\pi\sqrt{\frac{\bar{c}\,(\bar{L}_0-\bar{c}/24)}{6}}=S_{{\rm 3d}},
\end{equation}
using $L_0 - \frac{c}{24} = (M_{{\rm BTZ}}\ell_3 + J_{{\rm BTZ}})/2$ and $\bar{L}_0 - \frac{\bar{c}}{24}= (M_{{\rm BTZ}}\ell_3 - J_{{\rm BTZ}})/2$, with $M_{{\rm BTZ}},\,J_{{\rm BTZ}}$ given in \eqref{BTZ-MJ-BPS}.

We do not understand either in field theory or in gravitational terms, what the proper connection is between the description above and the one emerging in Kerr/CFT. We do want to emphasise that our approach did insist on taking a near horizon limit to study the IR properties of the system and keeping the latter on-shell. In this near-BPS regime, this forced us to lose part of the degrees of freedom responsible for the full black hole entropy. Both the Kerr/CFT central charge \eqref{eq:nearbpskerr} and our near horizon analysis do suggest the potential existence of a 2d CFT which is {\it not} intrinsically localised in the horizon. Similar observations have been made using low energy probes in (non-)extremal black holes \cite{More-Andy,Cvetic:2011hp}, without explicitly relying on the near horizon geometry.\footnote{If we start with a generic BPS or extremal black hole, take the near horizon limit first (as is done in \cite{Hartman:2008pb,Chow:2008dp}) and then take the nearly vanishing horizon limit, as we have done in this section, we obtain a different geometry than when we first take the near EVH black hole limit and then take the near horizon limit, as we did in sections \ref{sect:nearbps} and \ref{sect5}. Nonetheless, despite of having different geometries, as we have shown the entropies obtained in these two cases are equal. \label{order-of-limits-issue}}

It would be very interesting to provide {\it any} technical evidence confirming the structures uncover in gravity, along the lines of \cite{deMello}, by which the sector \eqref{BMN-BPS} in ${\cal N}=4$ SYM may be equivalently described by a (non-) chiral CFT whose degrees of freedom should be the open strings stretched between giant gravitons.

\subsection{Non-BPS extremal discussion}
\label{sec:nonbpskerr}

The purpose of this section is to extend the previous discussion to the near-extremal non-BPS limit \eqref{eq:FTcft}. Consider first
the limiting behaviour of the three Frolov-Thorne temperatures \eqref{jj21}
\begin{equation}
  T_1 = \frac{\epsilon}{\pi}\frac{\sqrt{\hat{q}_1}}{\hat{r}_+}, \quad \quad
  T_i \sim \frac{1}{\epsilon}, \quad i=2,3
\end{equation}
Since the black hole entropy scales like $S\sim N^2\epsilon$, central charges also have two distinctive behaviours
\begin{equation}
  c_1\sim N^2\,, \quad \quad c_i\sim \left(N\epsilon\right)^2\,\quad i=2,3
\end{equation}
For finite $N$, $c_i$ $i=2,3$ scale to zero, indicating the physical degrees of freedom scale to zero,
whereas $c_1$ remains finite.

When $N^2\epsilon\sim 1$, to keep the entropy finite, the same conclusion holds for $c_i$ : $c_i\sim\epsilon\to 0$, whereas $c_1$ diverges. This suggests two of the Kerr/CFT descriptions are breaking down, as expected, whereas the third one corresponds to a double scaling limit in which the CFT gap is scaled to zero, while one scales the vacuum energy $L_0-c_1/24\sim\epsilon$ to zero. This is precisely the set-up described in \cite{EVH-BTZ}, but now embedded in the AdS/CFT correspondence and explicitly implemented by an $N\to\infty$ limit.

Let us explore the relevant Kerr/CFT description more closely. Notice $T_1$ is related to the 2d chiral CFT temperature $T_L$ in \eqref{eq:cfttemp} by
\begin{equation}
  T_1=\frac{\epsilon}{\pi}\frac{h\hat{r}_+}{\sqrt{q_2q_3}}=\epsilon T_L = \epsilon\, \frac{y_+}{\pi\ell_3},
\end{equation}
where we used $r_+=\epsilon\hat{r}_+$ and the identity
\begin{equation}
  \hat{r}_+ = \frac{(q_2q_3)^{1/4}}{L}\,y_+ \quad \quad {\rm with} \quad \quad y_+^2 = L^2\frac{\sqrt{\hat{q}_1}}{h},
\end{equation}
encoding the extremality condition $y_+=y_-$, equivalent from \eqref{y-pm}, to $\hat{M} = 2\frac{h}{L^2}\sqrt{\hat{q}_1q_2q_3}$.

Comparing the central charge $c_1$ with $c_{{\rm AdS}_3}$ computed in the spacetime superconformal algebra associated to the near horizon pinching geometry \eqref{eq:nonbpsmetric}, one finds\footnote{One must reinsert $G_5=\pi L^3/(2N^2)$ into the expressions written in  \cite{Lu:2009gj}, where it was set to one, to reproduce this result.}
\begin{equation}
c_1= \frac{3\pi}{2G_5}\frac{q_2q_3}{h} = \frac{c_{{\rm AdS}_3}}{\epsilon}=c_{{\rm CFT}}.
\end{equation}
Thus, the Kerr/CFT central charge equals the central charge of the parent 2d CFT in \eqref{eq:parentcft}, that is the Brown-Henneaux central charge if we did not have the pinching.

\paragraph{Interpretation:} The Kerr/CFT central charge $c_1$ is {\it finite} before scaling Newton's constant. Due to the regime of charges tested in the (near-)EVH limit, there is an AdS${}_3$ throat emerging in the near horizon gravitational description that allows us to identify this CFT as the parent 2d CFT alluded to in section \ref{sec:IRnonbps}. The pinching orbifold is a {\it thermal} state in this parent CFT exploring very low energies, $L_0-c/24\to N^2\epsilon^2$ at low temperatures $T_1\sim \epsilon$. The spacetime conformal algebra perspective discussed in \cite{Martinec:2001cf} and mentioned in section \ref{sec:IRnonbps} corresponds to
the `long string sector' of this parent CFT.

This picture is basically the same as the one advocated in \cite{EVH-BTZ} for a similar situation in a much simpler setting of EVH BTZ black hole (i.e. massless BTZ): Taking the near horizon limit first and then going to near-EVH region, if we did not scale the central charge (or $N$), one would have ended up with the null self-dual orbifold. However, if we took the near-EVH limit first and then focused on the near horizon we end up with a pinching AdS$_3$ orbifold. One may start with some (parent) CFT with finite central charge and explores excitations with very low energy. Taking the $N\to\infty$ limit enables us to keep some non-trivial dynamics by considering the double scaling limit $L_0-c/24\to 0$ and $c\to\infty$, keeping the entropy {\it finite}.

\paragraph{Comment on the null self-dual orbifold:} In \eqref{eq:nonbpsnull}, we identified a different near horizon limit giving rise to a different local AdS${}_3$ throat, corresponding to the null self-dual orbifold. It is easy to see the central charge describing this throat is the same as in the previous discussion. The difference in scaling in the time coordinate and the absence of pinching are consistent with the chiral nature of the dual CFT description,\footnote{The latter may be understood as the DLCQ limit of a non-chiral 2d  CFT \cite{us,Balasubramanian:2003kq}.} which is identified with the one emerging in the limiting Kerr/CFT description. In this case, though, the absence of pinching suggests the surviving chiral sector is {\it not} decoupled.
It is worth stressing the appearance of non-trivial gauge fields in the transverse dimensions. Note also the discussion in footnote \ref{order-of-limits-issue}. These are reminiscent of the deformations recently discussed in \cite{ElShowk:2011cm}. It would be interesting to understand the physics of these.

\section{Summary and outlook}

In this work we continued studying aspects Extremal Vanishing Horizon (EVH) black holes in the family of static R-charged AdS$_5$ black holes, envisioning that a generic black hole in this family can be understood  as excitations above the EVH black hole. Generic (non-BPS) EVH black holes are defined as black holes with $A,\ T, G_N\to 0$ while the ratios $A/T$ and $A/G_N$  remain finite, where $A$ is the horizon area, $T$ is the Hawking temperature and $G_N$ is the Newton constant (BPS EVH black holes have some other subtleties we discussed in detail, see also below). We showed that in the near horizon limit of a generic EVH black hole one obtains an approximate AdS$_3$ throat. This AdS$_3$ throat for a near-EVH black hole turns to a BTZ black hole. In the case of EVH KK black hole it was shown that the near horizon limit is indeed a decoupling limit \cite{SheikhJabbaria:2011gc}. Unfortunately, we have not shown that the near horizon limit for the case of R-charged EVH black holes  is a decoupling limit.

We discussed that R-charged AdS$_5$ EVH black holes can be BPS or non-BPS and that the near horizon limit is only well-defined in the 10d uplift of the black hole. In the case of BPS EVH black holes taking the near horizon limit we were forced to also blow up  a four dimensional part ${\cal M}_4$ of  the 10d geometry, and hence it is appropriate to consider ``density'' of physical quantities over ${\cal M}_4$ and demand these densities to remain finite in the near horizon limit. In the non-BPS case
the 4d part of the geometry remains of finite volume while the AdS$_3$ throat becomes a pinching orbifold of AdS$_3$, a feature seemingly generic to all non-BPS EVH black holes. We discussed the possibility to resolve the pinching orbifold using large N and large central charge limits.

The appearance of an approximate near horizon AdS$_3$ throat motivated the EVH/CFT proposal \cite{SheikhJabbaria:2011gc}: near-EVH black holes or low energy excitations around an EVH black hole is described by a subsector of a 2d CFT. Moreover, when dealing with asymptotic AdS$_5$ black holes, there is also a UV dual CFT (${\cal N}=4$ SYM in this case). Based on the gravity picture we then proposed a relation between the IR and UV dual CFTs, in which the IR CFT is related to a BMN-like sector of the UV CFT. Exploring and establishing this proposal further is an open interesting question. In particular, in the non-BPS case, the IR physics is not governed by ground states, and it is therefore difficult to see how one could obtain an IR Hamiltonian which is not unbounded from below.

All emergent IR CFTs, whether related to AdS${}_3$ or AdS${}_2$,
share various common features. One such feature is that the
diffeomorphisms and associated Virasoro algebras that emerge involve
in some non-trivial way an $S^1$, which can be either an $S^1$ in
space on which the field theory lives (for rotating black holes) or
an internal $S^1$ (for R-charged black holes). Understanding the
relevant Virasoro algebras in the latter case requires one to
construct ``diffeomorphisms in R-charge space" directly in the field
theory. Though a difficult problem, finding these could provide a
key towards unraveling the way in which conformal symmetry emerges
in field theory. To give some idea of the sort of structure we might be looking for,
consider a 2d CFT with $N=(2,2)$ supersymmetry. Such theories have a
spectral flow symmetry, and let's denote the operator which spectral
flows by $n$ units (so it maps the NS and R sector into themselves)
by $S_n$. If we bosonize the $\U(1)$ current in the $\cN=(2,2)$
algebra and call the resulting boson $\chi$, then $S_n$ is crudely
speaking something like multiplication by $e^{i n\chi}$. Since the
zero modes of the current $J_0$ is approximately $J_0 \sim
\partial_{\chi}$, the operators $d_n\equiv S_n J_0$ resemble
standard ${\rm diff}\,S^1$ generators in the $\chi$-direction.
Indeed, one finds the Virasoro algebra $[d_n,d_m]=\frac{c}{3} (m-n)
d_{m+n}$. We leave a further examination of these operators, and
possible generalizations to other field theories to future work.

We also discussed how the EVH/CFT and Kerr/CFT proposals are related to each other along the lines of discussions in \cite{us,EVH-BTZ,terashima,recent-evh,terashima-2}: The chiral CFT appearing in Kerr/CFT in the near-EVH locus in the parameter space corresponds to the DLCQ of the 2d CFT whose pinching orbifold limit appears in the EVH/CFT. This, together with our discussions here, can potentially be used to identify the microscopic degrees of freedom of the dual 2d CFT and it would be interesting to pursue this further.

The EVH black holes are not limited to static ones and can be stationary, e.g. the one considered in  \cite{SheikhJabbaria:2011gc}. Within the class of asymptotic AdS$_5$ black holes we have a more general family of EVH black holes which involve rotation as well as R-charge. This class of charged-rotating AdS$_5$ EVH black holes will be studied in a future publication \cite{progress}.

\section*{Acknowledgements}

We would like to thank Vijay Balasubramanian for his collaboration at  different stages of this work and for many fruitful discussions over the years that took to type this note. We would also like to thank Geoffrey Comp\`ere, Noriaki Ogawa and Hossein Yavartanoo for useful discussions. The work of JdB was partly supported by the Foundation of Fundamental Research on Matter (FOM). The work of MJ and JS was partially supported by the Engineering and Physical Sciences Research Council (EPSRC) [grant number EP/G007985/1] and the Science and Technology Facilities Council (STFC) [grant number ST/J000329/1].


\appendix
\section{Near-horizon limits of singular Myers-Perry black
holes}\label{appendix}

As mentioned the appearance of AdS${}_3$ factors are not limited to asymptotic AdS$_5$ R-charged black holes and seems to be generic to any EVH black hole. One such example was 4d KK black hole discussed in \cite{SheikhJabbaria:2011gc}. Here we present another class which generalizes the Bardeen-Horowitz 5d Kerr EVH black hole \cite{bardeenhorowitz}. Let us start with revisiting the case of \cite{bardeenhorowitz}.
Their starting point is the 5d Myers-Perry
black hole \cite{myersperry} which generically come with angular momenta. These black holes are specified by three parameters, ADM
mass $m$ and two angular momentum parameters $a,b$. We take the
extremal limit for which $m=m(a,b)$, and then send one of the two
angular momenta, say the $b$ parameter, to zero in order to obtain a
vanishing horizon area, singular black hole. (In general the EVH
black holes are naked singularities.) The
resulting metric reads %
 \bea %
ds^2 & = & -d\tilde{t}^2 + \frac{a^2}{\rho^2}
(d\tilde{t}-a\sin^2\theta d\tilde{\phi})^2
 +(\tilde{r}^2+a^2)\sin^2\theta d\tilde{\phi}^2
\nonumber \\
& & + \tilde{r}^2 \cos^2\theta d\tilde{\psi}^2 + \rho^2 d\theta^2 +
\frac{\rho^2}{\tilde{r}^2} d\tilde{r}^2 ,%
\eea%
where $a$ parameterises the remaining angular momentum, and
$\rho^2=\tilde{r}^2 +a^2 \cos^2\theta$. The near-horizon    limit is
obtained by
\be\label{5d-EVH-Kerr-NH-limit}
\tilde{r}=\epsilon r\,,\quad \tilde{t}=t/\epsilon\,,\quad
\tilde{\phi}=\phi+\tilde{t}/a\,,
\quad\tilde{\psi}=\psi/\epsilon\,,
\ee%
and taking $\epsilon\rightarrow 0$. The result is \cite{bardeenhorowitz}%
\be\label{5d-Kerr-EVH-AdS}%
ds^2 = \cos^2 \theta \left[ -\frac{r^2}{a^2} dt^2
+\frac{a^2}{r^2} dr^2 + r^2 d\psi^2\right] + a^2 \left[ \cos^2\theta
d\theta^2 + \frac{\sin^2\theta}{\cos^2\theta} d\phi^2\right].%
\ee%
This has the structure of a warped AdS${}_3$, but the AdS${}_3$
collapses to zero size at $\theta=\pi/2$. In addition, the rescaling
of $\tilde{\psi}$ implies that $\psi$ has periodicity $2\pi
\epsilon$, in other words the $\psi$ circle shrinks to zero size in
the $\epsilon \rightarrow 0$ limit, which is the ``vanishing
periodicity'' pathology or pinching \ads3 orbifold problem which seems to be generic
behaviour for near-horizon limit of all \emph{non-BPS EVH black holes}.

One may extend the above limit to near-EVH 5d Kerr by ``perturbative addition'' of $b$, \ie by allowing
non-zero, but small $b$ parameter and small deviation of $m$
parameter from extremality. It is straightforward to show that with
appropriate scaling of $b$ and out-of-extremality (when they scale
like $\epsilon^2$) we obtain a geometry as \eqref{5d-Kerr-EVH-AdS}
but with AdS${}_3$ metric replaced with a BTZ geometry. The angular
momentum of the BTZ metric is measured by  $b/\epsilon^2$ parameter
while its mass by the out-of-extremality as $(m-m_{ext})/\epsilon^2$, such that if the original
near-EVH geometry is extremal  in the near-horizon    limit we
obtain an extremal BTZ. Note, however, that this BTZ is a ``pinching BTZ orbifold'', \emph{i.e.} it is a BTZ black hole
built upon the  the pinching \ads3 orbifold.

Inspired by the above near horizon limits and despite the fact that the geometry \eqref{5d-Kerr-EVH-AdS} has a (naked) singularity at $\theta=\pi/2$, one can show that upon the reduction of 5d
Einstein theory with the reduction ansatz
\be\label{reduction-5to3}
ds^2=\cos^2\theta
g_{\mu\nu}dx^\mu dx^\nu+R^2\left[ \cos^2\theta d\theta^2 +
\frac{\sin^2\theta}{\cos^2\theta} d\phi^2\right]\,,
\ee
we obtain a 3d Einstein gravity with cosmological constat $-\frac{1}{R^2}$. The 3d and 5d Newton constants are then related as
$G_3=\frac{G_5}{\pi R^2}$. The vacuum solutions to the 3d gravity theory obtained from this reduction are hence of the form of BTZ black holes, or other quotients of AdS$_3$ and the corresponding Brown-Henneaux central charge is $c=\frac{3\pi}{2}\frac{R^3}{G_5}$.

\paragraph{Higher dimensional Myers-Perry black holes:}
The above limit can be generalized to Myers-Perry
black holes in $d=2n+1$ dimensions. There one has $n$ angular
momentum parameters $a_i$. For the extremal black holes in this class, if one of $a_i$'s, say $a_n$, is zero we have an EVH black hole. In the near horizon limit of this EVH black hole  we obtain the metric%
\be\label{NH-general-Kerr-EVH}%
ds^2 = \mu_n^2 \left[ -\alpha r^2 dt^2
+\frac{1}{\alpha r^2} dr^2 + r^2 d\psi^2\right] + \sum_{i=1}^{n-1}
\left[ a_i^2 d\mu_i^2 + a_i^2 \mu_i^2\left(1+\frac{\mu_i^2}{\mu_n^2}
\right)d\phi_i^2\right] ,%
\ee%
where $\psi$ was the angle corresponding to the vanishing angular
momentum $a_n$, the $\phi_i$ are the angles corresponding to the
other angular momenta, and $\mu_i$ are coordinates on $S^{n-1}$ that
obey $\sum_{i=1}^n \mu_i^2=1$. The $\psi$ direction in the AdS$_3$ part has a vanishing periodicity and the parameter $\alpha$ is equal to $\alpha=\sum_{i=1}^{n-1}
\frac{1}{a_i^2}$. One may also consider near-EVH near-horizon     limit by a ``perturbative addition'' of $a_n$ and moving slightly away from extremality, as we described above for the 5d example. In this case one will get a (pinching) BTZ geometry instead of (pinching) AdS${}_3$.

The angle $\psi$ in \eqref{NH-general-Kerr-EVH} once more has vanishing periodicity (the 3d part is a pinching AdS$_3$). One can see from the explicit form of the solution
that taking a second angular momentum to zero does not lead to any
well-defined metric. In particular, there does not seem to be an
obvious generalization that leads to AdS${}_d$ metrics with $d>3$,
but it would be interesting to explore this in more detail.


\providecommand{\href}[2]{#2}\begingroup\raggedright

\endgroup
\end{document}